\documentclass[aps,pre,showpacs,showkeys,square,numbers,amssymb,amsmath,nobibnotes,nofootinbib]{revtex4}
\usepackage{docs}%
\usepackage{bm}%
\usepackage{amsmath}
\usepackage{amsfonts}
\usepackage{amssymb}
\usepackage{eucal}
\usepackage{verbatim}
\usepackage{times}

\usepackage{graphicx}
\usepackage[sort&compress]{natbib}

\def\Tr{\mathrm{Tr}}
\begin{document}

\title{ Probability distributions of Linear Statistics in Chaotic Cavities 
and associated phase transitions.}

\author{Pierpaolo Vivo}
\affiliation{Abdus Salam International Centre for
Theoretical Physics, Strada Costiera 11, 34151 Trieste, Italy}

\author{Satya N. Majumdar and Oriol Bohigas}
\affiliation{Laboratoire de Physique Th\'{e}orique et Mod\`{e}les
Statistiques (UMR 8626 du CNRS), Universit\'{e} Paris-Sud,
B\^{a}timent 100, 91405 Orsay Cedex, France}

\date{\today}

\begin{abstract}

We establish large deviation formulas for linear statistics on
the $N$ transmission eigenvalues $\{T_i\}$ of a chaotic cavity, in
the framework of Random Matrix Theory. Given any linear statistics
of interest $A=\sum_{i=1}^N a(T_i)$, the probability distribution
$\mathcal{P}_A(A,N)$ of $A$ generically satisfies the large deviation
formula $\lim_{N\to\infty}\left[-2\log\mathcal{P}_A(Nx,N)/\beta
  N^2\right]=\Psi_A(x)$,
where $\Psi_A(x)$ is a rate function that we compute explicitly in
many cases (conductance, shot noise, moments) and $\beta$ corresponds to different symmetry classes. Using these large
deviation expressions, it is possible to recover easily known
results and to produce new formulas, such as a closed form
expression for $v(n)=\lim_{N\to\infty}\mathrm{var}(\mathcal{T}_n)$
(where $\mathcal{T}_n=\sum_{i}T_i^n$) for arbitrary integer $n$.
The universal limit $v^\star=\lim_{n\to\infty} v(n)=1/2\pi\beta$
is also computed exactly. The distributions display a central Gaussian region flanked on both sides by non-Gaussian tails. At the junction of the two regimes,
weakly non-analytical points appear, a direct consequence of phase transitions in an associated Coulomb gas problem. Numerical
checks are also provided, which are in full agreement with our asymptotic results in both real and Laplace space even for moderately small $N$.
Part of the results have been 
announced in [P. Vivo, S.N. Majumdar and O. Bohigas, {\it Phys. Rev. Lett.} {\bf 101}, 216809 (2008)].
\end{abstract}


\maketitle

\section{Introduction}
We consider the statistics of quantum transport through a chaotic
cavity with $N_1=N_2=N\gg 1$ open channels in the two attached
leads. It is well established that the electrical current flowing
through a cavity of sub-micron dimensions presents time-dependent
fluctuations which persist down to zero temperature
\cite{beenakker} and are thus associated with the granularity of
the electron charge $e$. Among the characteristic features
observed in experiments, we can mention weak localization
\cite{chang}, universality in conductance fluctuations
\cite{marcus} and constant Fano factor \cite{oberholzer}.
In the Landauer-B\"{u}ttiker scattering approach
\cite{beenakker,landauer,buttikerPRL}, the wave function
coefficients of the incoming and outgoing electrons are related
through the unitary scattering matrix $S$ ($2N\times 2N$):
\begin{equation}\label{ScatteringMatrix S}
  S=
  \begin{pmatrix}
    r & t^\prime \\
    t & r^\prime
  \end{pmatrix}
\end{equation}
where the transmission ($t,t^\prime$) and reflection
$(r,r^\prime)$ blocks are $(N\times N)$ matrices encoding the
transmission and reflection coefficients among different channels.
Many quantities of interest for the experiments can be extracted
from the eigenvalues of the hermitian matrix $t t^\dagger$: for
example, the dimensionless conductance and the shot noise are
given respectively by $G=\Tr(t t^\dagger)$ \cite{landauer} and
$P=\Tr[t t^\dagger(1-t t^\dagger)]$ \cite{lesovik,ya}.

Random Matrix Theory (RMT) has been very successful in describing the statistics of universal fluctuations in such systems,
and complements the fruitful semiclassical approach \cite{richter,braun,berkolaiko,schanz,whitney}.
The simplest way to model the scattering matrix $S$ for the case
of chaotic dynamics is to assume that it is drawn from a suitable
ensemble of random matrices, with the overall constraint of
unitarity \cite{muttalib,stone,mellopereira,altshuler}. Through a
maximum entropy approach with the assumption of ballistic
point contacts \cite{beenakker}, one derives that the probability distribution of
$S$ should be uniform within the unitary group, i.e. $S$ belongs to
one of Dyson's Circular Ensembles \cite{Mehta,Dys:new}.

It is then a non trivial task to extract from this information the
joint probability density of the transmission eigenvalues $\{T_i\}$ of the
matrix $t t^\dagger $, from which the statistics of interesting experimental quantities could be in principle derived.
Fortunately, this can be done \cite{mellopereira,beenakker,forrcond,araujo} and
the expression reads:
\begin{equation}\label{jpd transmission First}
  P(T_1,\ldots,T_N)=A_N\prod_{j<k}|T_j-T_k|^\beta\prod_{i=1}^N
  T_i^{\beta/2-1}
\end{equation}
where the Dyson index $\beta$ characterizes different symmetry
classes ($\beta=1,2$ according to the presence or absence of
time-reversal symmetry and $\beta=4$ in case of spin-flip
symmetry). The eigenvalues $T_i$ are correlated random variables
between $0$ and $1$. The constant $A_N$ is explicitly known from
the celebrated Selberg's integral as:
\begin{equation}\label{A_N}
  A_N^{-1}=\prod_{j=0}^{N-1}\frac{\Gamma\left(1+\frac{\beta}{2}+j\frac{\beta}{2}\right)\Gamma\left(\frac{\beta}{2}(j+1)\right)
  \Gamma\left(1+j\frac{\beta}{2}\right)}{\Gamma\left(1+\frac{\beta}{2}\right)\Gamma\left(\frac{\beta}{2}+1+(N+j-1)\frac{\beta}{2}\right)}
\end{equation}

From \eqref{jpd transmission First}, in principle the
statistics of all observables of interest can be calculated. In this paper, we 
shall
focus on the following:
\begin{align}\label{List}
  G &= \sum_{i=1}^N T_i &\text{(conductance)}\\
  P &= \sum_{i=1}^N T_i(1-T_i) &\text{(shot noise)}\\
  \mathcal{T}_n &= \sum_{i=1}^N T_i^n &\text{(integer moments)}
\end{align}
although in principle any linear statistics\footnote{$A$ is a
linear statistics in that it does not contain products of different eigenvalues.
The function $a(x)$ may well depend non-linearly on $x$.}
$A=\sum_{i=1}^N a(T_i)$ can be tackled with the method described
below. It is worth mentioning that an increasing interest for the
moments and cumulants of the transmission eigenvalues can be
observed in recent theoretical \cite{blanter} and experimental
works \cite{lu}.

Many results are known for the average and the variance of the
above quantities, both for large $N$
\cite{beenakker,brouwer,novaes} and, very recently, also for a
fixed and finite number of channels $N_1,N_2$
\cite{novaes,vivovivo,savin,savin2}. In particular, a general formula for
the variance of any linear statistics  $A=\sum_{i=1}^N
a(\lambda_i)$ (where $T_i=(1+\lambda_i)^{-1}$) in the limit of
large number of open channels is known from Beenakker
\cite{beenakkerPRL}. However, at least for the case
$A=\mathcal{T}_n$ (integer moments), this formula is of little
practical use. The method we introduce below allows to obtain the
sought quantity
$v(n)=\lim_{N\to\infty}\mathrm{var}(\mathcal{T}_n)$ in a neat and
explicit way.

In contrast with the case of mean and variance, for which a wealth
of results are available (see e.g. \cite{beenakker,bulgakov} and references therein), much less is known for the \emph{full}
distribution of the quantities above: for the conductance, an
explicit expression was obtained for $N_1=N_2=1,2$
\cite{baranger,jalabert,garcia}, while more results are available
in the case of quasi one-dimensional wires \cite{muttalibwolfle}
and 3D insulators \cite{muttalib3D}. For the shot noise, the
distribution was known only for $N_1=N_2=1$ \cite{pedersen}. Very
recently, Sommers {\it et al.} \cite{sommers} announced two
formulas for the distribution of the conductance and the
shot noise, valid at arbitrary number of open channels and for any $\beta$, which are
based on Fourier expansions. Such results are then incorporated and expanded in \cite{savinnew}.
In \cite{Kanz} and \cite{Kanz2}, along with recursion formulas for the efficient computation of conductance and shot noise cumulants, an asymptotic analysis 
for the distribution functions of these quantities in the limit of many open channels was reported.
In a recent letter \cite{vivoPRL}, we announced the computation of the same asymptotics (in the form of large deviation expressions) 
using a Coulomb gas method and we pointed out a significant discrepancy with respect to the claims by Osipov and Kanzieper \cite{Kanz,Kanz2}.
We will discuss in detail such disagreement and the way to sort it out convincingly in subsection \ref{comp:sect} for the conductance case for $\beta=2$.

Given some recent experimental progresses \cite{hemmady}, which
made eventually possible to explore the full distribution for the
conductance (and not just its mean and variance), it is of great
interest to deepen our knowledge about distributions of other
quantities, whose experimental test may be soon within reach.

It is the purpose of the present paper to build on \cite{vivoPRL} and establish exact large
deviation formulas for the distribution of any linear statistics
on the transmission eigenvalues of a symmetric cavity with
$N_1=N_2=N\gg 1$ open channels. More precisely, for any linear
statistics $A$ whose probability density function (pdf) is denoted as
$\mathcal{P}_A(A,N)$, we have\footnote{Hereafter we will use the
concise notation
$\mathcal{P}(A,N)\approx\exp\left(-\frac{\beta}{2}N^2\Psi_A(A/N)\right)$
to mean exactly \eqref{LargeDeviationsLinearStatistics}.}:
\begin{equation}\label{LargeDeviationsLinearStatistics}
  \lim_{N\to\infty}\left[-\frac{2\log\mathcal{P}_A(A=Nx,N)}{\beta
  N^2}\right]=\Psi_A(x)
\end{equation}
where the large deviation function $\Psi_A(x)$ (usually called \emph{rate function}) is computed exactly for
conductance $(A=G)$, shot noise $(A=P)$ and integer moments
$(A=\mathcal{T}_n)$. The method can be extended to the case of
asymmetric cavities $N_1=\kappa N_2$, and
is based on a combination of the standard Coulomb gas analogy by Dyson and
functional methods recently exploited in \cite{DM} in the context
of Gaussian random matrices. This approach has been then
fruitfully applied to many other problems in statistical physics
\cite{vivomaj,BrayDean,FSW,FyodWilliams,nadal,vergassola,verg2,vivoindex}.

The paper is organized as follows. Section II provides a quick
summary of our main results. In Section III, we summarize 
the Coulomb gas method that can in principle be used to
obtain the rate function associated with
arbitrary linear statistics. In Section IV, V and VI, we use
this general method to obtain explicit results respectively
for the conductance, shot noise and integer moments.
Finally we conclude in Section VII with a summary and some
open problems.

\section{Summary of results}\label{summarysect}

\noindent {\bf Distribution of conductance $G$:} We obtain the following exact expression for the rate 
function in the case of conductance ($G$)  
\begin{align}
  \label{PsiGsummary} \Psi_G(x) &=
  \begin{cases}
  \frac{1}{2}-\log(4x) & \text{for}\quad 0\leq x\leq \frac{1}{4}\\
    8\left(x-\frac{1}{2}\right)^2 & \text{for}\quad \frac{1}{4}\leq x\leq \frac{3}{4} \\
\frac{1}{2}-\log[4(1-x)] & \text{for}\quad \frac{3}{4}\leq x\leq 1
 \end{cases}
\end{align}
The rate function has a quadratic form near its minimum at $x=1/2$ in
the range $ 1/4\le x\le 3/4$. 
Using \eqref{LargeDeviationsLinearStatistics}, it then follows that the
distribution $\mathcal{P}_G(G,N)$ has a Gaussian form close to its peak
\begin{widetext}
\begin{equation}\label{AbbreviatedLaw}
  \mathcal{P}_G(G,N)\approx\exp\left(-\frac{\beta}{2}N^2\Psi_G\left(\frac{G}{N}\right)\right)
  =\exp\left[-\frac{\beta}{2}N^2\cdot
  8\left(\frac{G}{N}-\frac{1}{2}\right)^2\right]=
  \exp\left[-\frac{1}{2(1/8\beta)}\left(G-\frac{N}{2}\right)^2\right]
\end{equation}
\end{widetext}
from which one easily reads off the 
mean and variance $\langle G\rangle=N/2$ and
$\mathrm{var}(G)=1/8\beta$, which agree with their known large $N$ values.
The fact that the variance becomes indepedent of $N$ for large $N$ is
referred to as the {\it universal conductance fluctuations}. This central
Gaussian regime is valid over the region $N/4\le G\le 3N/4$. Outside this
central zone, $\mathcal{P}_G(G,N)$ has non-Gaussian large deviation power law tails. 
Using \eqref{PsiGsummary} in \eqref{LargeDeviationsLinearStatistics}
near $x=0$ and $x=1$ , we get $\mathcal{P}_G(G,N)\sim G^{\beta N^2/2}$ (as $G\to 0$)
and $\mathcal{P}_G(G,N)\sim (N-G)^{\beta N^2/2}$ (as $G\to N$) which are
in agreement, to leading order in large $N$, with the exact far tails obtained in ~\cite{mello22,sommers,savinnew}.
It turns out that while the central Gaussian regime matches continuously with
the two side regimes, there is 
a weak singularity at the two
critical points $G=N/4$ and $G=3N/4$ (only the 3rd derivative is discontinuous).
One of our central results is to show that these two weak singularities 
in the conductance distribution arise due to two phase transitions in
the associated Coulomb gas problem where the average charge density
undergoes abrupt changes at these critical points. 
We note that an intermediate regime
with an exponential tail claimed in \cite{Kanz,Kanz2} does not appear in our solutions, and we will comment in detail about such discrepancy in subsection \ref{comp:sect}. \\

\noindent {\bf Distribution of shot noise $P$:} For the shot noise, the exact
rate function reads
\begin{align}
\label{PsiPsummary}\Psi_P(x) &=
  \begin{cases}
     \frac{1}{4}-2\log 2-\frac{1}{2}\log x & \!\!\!\!\text{for}\quad\!\!\!\! 0\leq x\leq \frac{1}{16}\\
    64\left(x-\frac{1}{8}\right)^2 & \!\!\!\!\text{for}\quad\!\!\!\!\frac{1}{16}\leq x\leq \frac{3}{16} \\
    \frac{1}{4}-2\log 2-\frac{1}{2}\log\left(\frac{1}{4}-x\right) & \!\!\!\!\text{for}\quad\!\!\!\! \frac{3}{16}\leq x\leq \frac{1}{4}
  \end{cases}
\end{align}
Thus the shot noise distribution $\mathcal{P}_P(P,N)$, in the large $N$ limit, also
has a central Gaussian regime over $N/16 \le P\le 3N/16$ where
\begin{widetext}
\begin{equation}\label{AbbreviatedLawNoise}
  \mathcal{P}_P(P,N)\approx\exp\left(-\frac{\beta}{2}N^2\Psi_P\left(\frac{P}{N}\right)\right)
  =\exp\left[-\frac{\beta}{2}N^2\cdot
  64\left(\frac{P}{N}-\frac{1}{8}\right)^2\right]=
  \exp\left[-\frac{1}{2(1/64\beta)}\left(P-\frac{N}{8}\right)^2\right]
\end{equation}
\end{widetext}
yielding $\langle P\rangle=N/8$ and $\mathrm{var}(P)=1/64\beta$. As in the case of conductance,
this central Gaussian regime is flanked on both sides by two non-Gaussian power law tails with
weak third order singularities at the transition points $P=N/16$ and $P=3N/16$. \\

\noindent {\bf Distribution of integer moments $\mathcal{T}_n$:} For general $n>1$, the
computation of the full rate function $\Psi_{\mathcal{T}_n}(x)$ 
and hence the large $N$ behavior of the distribution $\mathcal{P}_{\mathcal{T}_n}(\mathcal{T}_n,N)$ 
turns out to be rather cumbersome.
However, the distribution $\mathcal{P}_{\mathcal{T}_n}(\mathcal{T}_n,N)$
shares some common qualitative features with the conductance and the shot noise distributions.
For example, we show that for any $n>1$, there is a central Gaussian regime where
the rate function $\Psi_{\mathcal{T}_n}(x)$ has a quadratic form given exactly by:
\begin{equation}
\Psi_{\mathcal{T}_n}(x)=\frac{b_n}{2}\left[x-\frac{\Gamma(n+1/2)}{\sqrt{\pi}\Gamma(n+1)}\right]^2
\end{equation}
where $b_n =\frac{4\pi\Gamma(n)\Gamma(n+1)}{[\Gamma(n+1/2)]^2}$. This implies
a Gaussian peak in the distribution:
\begin{equation}\label{DecayPTn}
\mathcal{P}_{\mathcal{T}_n}(\mathcal{T}_n,N)\approx
\exp\left[-\frac{\beta}{2}N^2 \frac{b_n}{2}\left[\frac{\mathcal{T}_n}{N}-\frac{\Gamma(n+1/2)}{\sqrt{\pi}\Gamma(n+1)}\right]^2\right]
\end{equation}
\begin{figure}[htb]
\includegraphics[bb =14 14 389 246, width=0.4\textwidth]{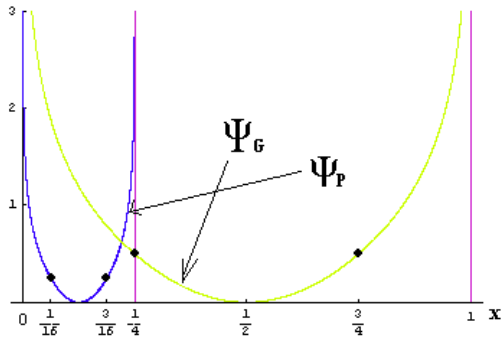}
\caption{(Color online). $\Psi_G(x)$ and $\Psi_P(x)$ as in \eqref{PsiGsummary} and \eqref{PsiPsummary}.
The black dots highlight the critical points on each curve.\label{PsiGfig}}
\end{figure}
from which one easily reads off the mean and variance of integer moments:
\begin{align}
\langle\mathcal{T}_n\rangle &=\frac{N\Gamma(n+1/2)}{\sqrt{\pi}\Gamma(n+1)}\\
\mathrm{var}(\mathcal{T}_n) &=\frac{2}{\beta b_n}=\frac{[\Gamma(n+1/2)]^2}{2\beta\pi\Gamma(n)\Gamma(n+1)}
 \end{align}
To the best of our knowledge, the latter result, together with its universal asymptotic value 
$\lim_{n\to\infty}\mathrm{var}(\mathcal{T}_n)=(2\beta\pi)^{-1}$, has not been reported previously in the literature.

As in the case of conductance and shot noise, there are singular points separating the central Gaussian regime 
and the non-Gaussian tails. However, the situation is more complicated for $n>1$. Unlike in the conductance or 
the shot noise case where there are only two phase transitions separating three regimes, there are more 
regimes for $n>1$. We analyse here in detail the case $n=2$ where we show that there are actually three 
singular points separating $4$ regimes. This is again a consequence of phase transitions in the associated 
Coulomb gas problems where the average charge density undergoes abrupt changes at these critical points. 
It would be interesting
to see how many such phase transitions occur for $n>2$. However,
we do not have any simple way to predict this and this remains
an interesting open problem. 
The 
tails of the distribution for the integer moments case can be computed in principle for general $n>1$, and the 
formalism is developed in section \ref{distmom}, but for clarity we will focus mostly on the case $n=2$.
The left tail has again a power-law decay for any $n>1$.

\section{ Summary of the Coulomb gas method}\label{summarycoulomb}

It is useful to summarize briefly the method we use to compute
the distribution of linear statistics in the large $N$ limit.
Given a linear statistics of interest $A=\sum_{i=1}^N a(T_i)$, its probability
density function (pdf) $\mathcal{P}_A(A,N)$, using the joint pdf of $T_i$'s in
Eq. (\ref{jpd transmission First}), is given by
\begin{widetext}
\begin{equation}\label{PA(A,N)}
  \mathcal{P}_A(A,N)=A_N\int_0^1\cdots\int_0^1 dT_1\cdots dT_N
\exp\left(\frac{\beta}{2}\sum_{j\neq
k}\log|T_j-T_k|+\left(\frac{\beta}{2}-1\right)\sum_{i=1}^N \log
T_i\right)\delta\left(\sum_{i=1}^N a(T_i)-A\right).
\end{equation}
\end{widetext}
The main idea then is to work with its Laplace transform
\begin{equation}\label{Laplacez}
\mathbf{F}_N\left(z;A\right)=\int_0^\infty dA\ \mathcal{P}_A(A,N)e^{-\frac{\beta}{2}zA}
\end{equation}
where the variable $z$ takes in principle complex values and its dependence on $N$ is at present unspecified.

Suppose one is able to compute the following limit:
\begin{equation}\label{limitJP}
J_A(p)=-\frac{2}{\beta}\lim_{N\to\infty}\frac{\log \mathbf{F}_N\left(N^\alpha p;A\right)}{N^2}
\end{equation}
and such limit is finite and nonzero for a certain \emph{speed} $\alpha$ and for $p\in\mathbb{R}$ such that $p\sim\mathcal{O}(1)$ for $N\to\infty$.
It is then a classical result in large deviation theory (G\"{a}rtner-Ellis theorem, see e.g. \cite{touchette} Appendix C) that the following finite nonzero limit exists:
\begin{equation}\label{limitrateGE}
\Psi_A(x)=-\frac{2}{\beta}\lim_{N\to\infty}\frac{\log \mathcal{P}_A\left(N^\alpha x,N\right)}{N^2}
\end{equation}
and $\Psi_A(x)$ [the \emph{rate function}] is given by the inverse Legendre transform of $J_A(p)$:
\begin{equation}
\Psi_A(x) = {\rm max}_p\left[J_A(p)-p\, x\right].
\label{inverse1}
\end{equation}
Note that:
\begin{enumerate}
\item There is no need to consider \emph{complex} values for $z$ (and thus for $p$) in \eqref{Laplacez}, as only real values matter for obtaining the rate function.
As $\mathcal{P}_A(A,N)$ has generically a compact support, negative values for $p$ are also allowed.
\item Setting the appropriate speed $\alpha$ is evidently \emph{crucial} for obtaining a finite nonzero limit in \eqref{limitJP}. There is no freedom in here. Conversely, \emph{any} speed is equally
good when extracting the \emph{cumulants} out of the Laplace transform \eqref{Laplacez} $(z=N^\alpha p)$ through the formula (set $\beta=2$ for simplicity):
\begin{equation}\label{cum1}
\kappa_\ell(A)=\left(\frac{-1}{N^\alpha}\right)^\ell \frac{\partial^\ell}{\partial p^\ell}\log \mathbf{F}_N(N^\alpha p;A)\Big|_{p=0}
\end{equation}
as no limiting procedure is performed in \eqref{cum1}.
\item The rate function $\Psi_A(x)$ encodes the full information about the probability distribution in the limit of infinitely many open channels. However, our numerical simulations
confirm that it also gives a fairly accurate description of such distribution for rather small $N$.
\end{enumerate}
What is then the correct speed $\alpha$ for the linear statistics considered here? It is quite easy to argue that $\alpha$ must be set equal to $1$. The reason is best
understood by taking the conductance $G=\sum_{i=1}^N T_i$ for $\beta=2$ as an example. By very general arguments we expect the large $N$
behavior of $\mathcal{P}_G(G,N)$ to scale as $\mathcal{P}_G(G,N)\sim e^{-N^2 \Psi_G(G/N)}$ for large $N$. Clearly, the two exponentials (the one coming from $\mathcal{P}_G(G,N)$ and the other coming from the Laplace measure)
must be of the same order in $N$ to guarantee a meaningful saddle point contribution, and since $G\sim N$ for large $N$, clearly $z\sim N$ as well.

After setting the proper speed, we get in full generality:

\begin{widetext}
\begin{equation}\label{LaplaceTransformA}
  \underbrace{\int_0^\infty \mathcal{P}_A(A,N)e^{-\frac{\beta}{2}NpA}dA}_{\mathbf{F}_N (N p;A)}=
  A_N\int_0^1\!\!\!\!\cdots\!\!\!\int_0^1 dT_1\cdots dT_N
\exp\left(\frac{\beta}{2}\sum_{j\neq
k}\log|T_j-T_k|+\left(\frac{\beta}{2}-1\right)\sum_{i=1}^N \log
T_i-\frac{\beta}{2}pN\sum_{i=1}^N a(T_i)\right).
\end{equation}
\end{widetext}
We can write the exponential as, $\exp\left[-\beta E(\{T_i\})\right]$
with $E(\{T_i\})= -(1/2) \sum_{j\ne k} \log |T_j-T_k| +\sum_i V(T_i)$
where $V(T)= (1/\beta-1/2)\, \log(T) +pN a(T)/2$. 
This representation provides a natural
Coulomb gas interpretation. We can identify $T_i$'s as the
coordinates of the charges of a $2$-d Coulomb gas that lives on
a one dimensional real segment $[0,1]$. The charges repel each other
via the $2$-d logarithmic Coulomb potential and in addition, 
they sit in an external potential $V(T)$. 
Note that the Laplace parameter $p$ appears explicitly
in the external potential $V(T)$.
Then $E$ is the energy of this Coulomb gas.
Thus one can write
the Laplace transform as the ratio of two partition functions
\begin{equation}
 \mathbf{F}_N\left(N p;A\right)=\int_0^\infty \mathcal{P}_A(A,N)e^{-\frac{\beta}{2}NpA}dA=\frac{Z_p(N)}{Z_0(N)}
\label{ratio1}
\end{equation}
where $Z_p(N)$ is precisely the multiple integral on the rhs of Eq. (\ref{LaplaceTransformA})
and $Z_0(N)= 1/A_N$ (which simply follows by putting $p=0$ in Eq. (\ref{LaplaceTransformA})
and using the fact that the pdf $\mathcal{P}_A(A,N)$ is normalized to unity).

The next step is to evaluate this partition function $Z_p(N)$ 
of the Coulomb gas in the large $N$ limit.
This procedure for the large $N$ calculation was originally introduced by Dyson~\cite{Dys:new}
and has recently been used in the context of the largest eigenvalue distribution of
Gaussian~\cite{DM} and Wishart random matrices~\cite{vivomaj} and also in other related problems
of counting stationary points in random Gaussian landscapes~\cite{BrayDean}.   
There are two basic steps involved. 
The first step is a coarse-graining procedure where one sums over (partial 
tracing) all
microscopic
configurations of $T_i$'s compatible with a fixed charge density function $\varrho_p(T)=N^{-1}\sum_i
\delta(T-T_i)$ and the second step consists in performing a functional integral
over all possible positive charge densities $\varrho_p(T)$ that are normalized to unity.
Finally the functional integral is carried out in the large $N$ limit by the saddle point
method. 

Following this general procedure summarized in ~\cite{DM},
the resulting functional integral, to leading order in large $N$, becomes:
\begin{equation}
Z_p(N) \propto \int\mathcal{D}[\varrho_p]e^{-\frac{\beta}{2}N^2 S[\varrho_p]}
\label{pf2}
\end{equation}
where the action is given by
\begin{align}\label{action1}
 \nonumber S[\varrho_p] &=p\int_0^1 \varrho_p(T)\, a(T) dT+B\left[\int_0^1 \varrho_p(T)dT-1\right]\\
 &-\int_0^1\int_0^1 dTdT^\prime
  \varrho_p(T)\varrho_p(T^\prime)\log|T-T^\prime|.
\end{align}
Here $B$ is a Lagrange multiplier enforcing the normalization of
$\varrho_p(T)$. In the large $N$ limit, the functional integral in \eqref{pf2} is particularly suitable to be evaluated by the 
saddle 
point method\footnote{Note that such a nice feature is a direct consequence of having employed the \emph{correct} speed $\alpha=1$, i.e. of having scaled the Laplace parameter $z$ with $N$
in the correct way.}, i.e.,
one finds the solution $\varrho_p^\star(T)$ (the equilibrium charge density that minimizes
the action or the free energy) from the stationarity condition $\delta
S[\varrho_p]/\delta\varrho_p=0$ which leads to an
integral equation
\begin{equation}
V_{\rm ext}(T)= p\, a(T) + B = 2\int_0^1 \varrho_p^\star (T^\prime)\, \log|T-T^\prime|\, dT' 
\label{inteq1}
\end{equation}
where $V_{\rm ext}(T)=pa(T)+B$ is termed as external potential.
Differentiating once with respect to $T$ leads to a singular integral equation
\begin{equation}
\frac{p}{2}\, a'(T) ={\rm
Pr}\int_0^1\frac{\varrho_p^\star(T^\prime)}{T-T^\prime}dT^\prime
\label{sing1}
\end{equation}
where ${\rm Pr}$ denotes the principal part and $a'(T)= da/dT$.

Assuming one can
solve \eqref{sing1} for $\varrho_p^\star$, one next evaluates the
action $S[\varrho_p]$ in \eqref{action1} at the stationary
solution $\varrho_p^\star$ and then takes the ratio in
\eqref{ratio1} to get (upon comparison with \eqref{limitJP}):
\begin{equation}
\label{AsymptoticDecay}
\underbrace{\int_0^\infty \mathcal{P}_A(A,N)e^{-\frac{\beta}{2}NpA}dA}_{\mathbf{F}_N(Np;A)}\approx e^{-\frac{\beta}{2}N^2
[\overbrace{S[\varrho_p^\star]-S[\varrho_0^\star]}^{J_A(p)}]}.
\end{equation}
Inverting the Laplace transform gives the main asymptotic result $
\mathcal{P}(A,N)\approx \exp\left(-\frac{\beta}{2}N^2
\Psi_A\left(\frac{A}{N}\right)\right) $ where the rate function
is the inverse Legendre transform (see \eqref{inverse1}),
\begin{equation}
\Psi_A(x)=\max_p[-x\, p + J_A(p)]
\label{legendre1}
\end{equation}
with $J_A(p)$ given by the {\em free energy difference}
as in \eqref{AsymptoticDecay}.

To summarize, given any linear statistics $a(T)$, the steps are: (i) solve the singular
integral equation (\ref{sing1}) for the density $\varrho_p^\star(T)$ (ii) evaluate the action
$S[\varrho_p^\star]$ in \eqref{action1} (iii) evaluate $J_A(p)= S[\varrho_p^\star]-S[\varrho_0^\star]$
and finally (iv) use $J_A(p)$ in \eqref{legendre1}, maximize the rhs to evaluate the
rate function $\Psi_A(x)$. We will see later that all these steps can be carried
out fully and explicitly when $A=G$ (conductance) and $A=P$ (shot noise) and 
partially when $A=\mathcal{T}_n$ (integer moments).
 
The important first step is to find the explicit solution of the singular
integral equation \eqref{sing1}. Note that this equation is of the Poisson form
and it is, in some sense, an inverse electrostatic problem: given the potential
$p\,a'(T)$, we need to find the charge density $\varrho_p^\star(T)$. 
To proceed, we recall a theorem due to Tricomi~\cite{Tricomi1} concerning
the general solution to singular integral equations of the form
\begin{equation}
\label{intequation1}
g(x)=\mathrm{Pr}\int_a^b\frac{f(x^\prime)}{x-x^\prime}dx^\prime
\end{equation}
where $g(x)$ is given and one needs to find $f(x)$ which has only
a single support over the interval $[a,b]$ with $a<b$ (the lower edge $a$ 
should not be confused with the linear statistics function $a(T)$).
The solution $f(x)$, with a single support over $[a,b]$ can be found
explicitly~\cite{Tricomi1}
\begin{widetext}
\begin{equation}
\label{formulatricomi}
f(x)=-\frac{1}{\pi^2\sqrt{(b-x)(x-a)}}
\left[\mathrm{Pr}\int_a^b\frac{\sqrt{(b-x^\prime)(x^\prime-a)}}
{x-x^\prime}g(x^\prime)dx^\prime+B_1\right]
\end{equation}
\end{widetext}
where $B_1$ is an arbitrary constant. 

In our case, $g(x) = p\, a'(x)/2$ and provided we assume that the charge density has
a single support over $[a,b]$ with $a < b$, we can in principle 
use this solution \eqref{formulatricomi}.
However, if the solution happens to have a disconnected support
one cannot use this formula directly. Whether the solution has a single or disconnected
support depends, of course, on the function $a(T)$. We will see that
indeed for the case of conductance ($a(T)=T$), the solution has
a single support and one can use \eqref{formulatricomi} directly.
The edges $a$ and $b$
in that case are determined self-consistently as explained in Section IV.
On the other hand, for the shot noise ($a(T)= T(1-T)$) and for integer
moments with $n>1$ ($a(T) = T^n$), it turns out that for certain values
of the parameter $p$, the solution has a disconnected support. In that
case, one cannot use \eqref{formulatricomi} directly. However,
we will see later that one can still obtain the solution explicitly
by an indirect application of \eqref{formulatricomi}.    

A very interesting feature of \eqref{sing1} is that, depending on the value of the Laplace parameter $p$, 
the fluid of charged particles 
undergoes a series of real phase transitions in Laplace space, i.e., as one varies the Laplace parameter $p$,
there are certain critical values of $p$ at which the solution $\varrho_p^\star(T)$ abruptly changes its form. 
As a consequence, the rate function, related to the Laplace transform via the Legendre transform 
\eqref{legendre1}, also undergoes a change of behavior as one varies its argument at the corresponding 
critical points. The rate function is continuous at these critical points but it has 
weak non-analiticities (characterized by a discontinuous third derivative). 

\begin{figure}[htb]
\begin{center}
\includegraphics[bb =14 14 527 399,totalheight=0.27\textheight]{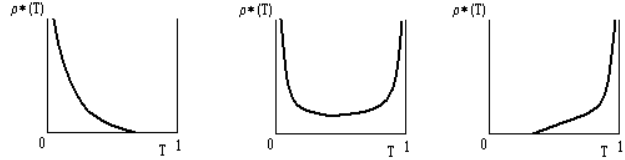}
\caption{Phases of the density of transmission eigenvalues for the
conductance case.\label{PhaseCond}}
\end{center}
\end{figure}

As an example, we consider the case of the conductance ($a(T)=T$)
(see section \ref{condsect} for details). In fig. \ref{PhaseCond},
we plot schematically the saddle point density
$\varrho_p^{\star}(T)$ (solution of \eqref{sing1} with $a(T)=T$), for three
different intervals on the real $p$ line. We will see in the next section
that there are three possible solutions valid respectively for $p\geq 4$,
$-4\le p\le 4$ and $p\le -4$.

\begin{itemize}
  \item when $p\geq 4$, the external potential $V_{\rm ext}(T)= p\, T+B$
is strong enough (compared
to the logarithmic repulsion) to keep the fluid particles confined
between the hard wall at $T=0$ and a point $L_1=4/p$. The gas
particles accumulate towards $T\to 0^{+}$, where the density
develops an inverse square root divergence, while
$\varrho_p^\star(L_1)=0$. This situation is depicted in the left panel
of fig. \ref{PhaseCond};
\item when $p$ hits the critical value $p^{(+)}=4$ from above, the density
profile changes abruptly. The external potential
$V_{\mathrm{ext}}$ is no longer overcoming the logarithmic
repulsion, so the fluid
  particles spread over the whole support $(0,1)$, and the density
  generically exhibits an inverse square root divergence at both
  endpoints ($T\to 0^{+}$ and $T\to 1^{-}$). This phase keeps
  holding for all the values of $p$ down to the second critical
  point $p^{(-)}=-4$ (see the second panel in fig. \ref{PhaseCond}), when the negative slope
  of the potential is so steep that the particles can no longer spread over the whole support $(0,1)$,
  but prefer to be located near the right hard edge at $T=1$.
  \item In the third phase $(p<-4)$, the fluid particles are
  pushed away from the origin and accumulate towards the right
  hard wall at $T\to 1$ (see the rightmost panel in fig.
  \ref{PhaseCond}). The density thus vanishes below the point
  $L_2=1-4/|p|$.
\end{itemize}

It is worth mentioning that such phase transitions in the solutions
of integral equations have been observed recently in other systems
that also allow similar Coulomb gas representations. These include
bipartite quantum entanglement problem~\cite{facchi}, nonintersecting
fluctuating interfaces in presence of a substrate~\cite{nadal}
and also multiple input multiple output (MIMO) channels~\cite{moustakas}.

\section{Distribution of the conductance}\label{condsect}

We start with the simplest case of linear statistics, namely the
conductance $G=\sum_{i=1}^N T_i$. Thus in this case $a(T)=T$ is
simply a linear function. Substituting $a(T)=T$ in \eqref{inteq1} 
gives
\begin{equation}\label{StationaryPhase}
V_{\rm ext}(T)=  p\,T + B=2\int_0^1 \varrho_p^\star(T^\prime)\log|T-T^\prime|dT^\prime
\end{equation}
and then \eqref{sing1} becomes
\begin{equation}\label{Tricomisol}
  \frac{p}{2}=\mathrm{Pr}\int_0^1\frac{\varrho_p^\star(T^\prime)}{T-T^\prime}dT^\prime
\end{equation}

We have then to find the solution to \eqref{Tricomisol}. Once this solution
$\varrho_p^\star(T)$ is found, we can evaluate the action
$S[\varrho_p^\star]$ at the saddle point in the following way. 
Multiplying \eqref{StationaryPhase} by $\varrho_p^\star(T)$ and integrating (using the normalization $\int_0^1 \varrho_p^\star(T) dT =1$) gives
\begin{align}
\nonumber p \int_0^1 T \varrho_p^\star(T)\, dT + B &= 2 \int_0^1\int_0^1 
\varrho_p^\star(T)\,\varrho_p^\star(T')\times\\
&\times\log|T-T'|\,dT\, dT'.
\label{step1}
\end{align}
Next we use this result to replace the double integral term in the action in \eqref{action1}
(with $a(T)=T$) to get
\begin{equation} 
S[\varrho_p^\star]= \frac{p}{2} \int_0^1 \varrho_p^*(T) T dT - \frac{B}{2}.
\label{step2}
\end{equation}
The yet unknown constant $B$ is determined from \eqref{StationaryPhase} upon using
the explicit solution, once found.
 
To find the solution to \eqref{Tricomisol} explicitly we will use the general
Tricomi formula in \eqref{formulatricomi} assuming a single support
over $[a,b]$. The edges $a$ and $b$ will be determined self-consistently.
Physically, we
can foresee three possible forms for the density
$\varrho_p^\star(T)$ according to the strength and sign of the
external potential $V_{\rm ext}(T)= p\,T+B$ on the left hand side (lhs) of \eqref{StationaryPhase}.
\begin{enumerate}
  \item For large and positive $p$, the fluid particles (transmission eigenvalues) 
will feel a strong confining potential
which keeps them close to the left hard edge $T\to 0+$. 
Thus, $\varrho_p^\star(T)$ will have a
support $[0,L_1]$, with $0<L_1<1$.
  \item For intermediate values of $p$, the particles will spread over the full range $[0,1]$.
  \item For large and negative $p$, the fluid particles will be pushed towards the right edge and 
the support of $\varrho_p^\star(T)$
will be over $[L_2,1]$, with $0<L_2<1$.
\end{enumerate}

These three cases will correspond to different solutions for the
Tricomi equation \eqref{Tricomisol} above, and the positivity
constraint for the obtained densities will fix the range of
variability for $p$ in each case. 
Once a solution
$\varrho_p^\star(T)$ for each case (different ranges for $p$) is
obtained, we can then use \eqref{step2} to evaluate the action
at the saddle point.

Let us consider the three cases discussed above separately.

\subsection{Large $p$: support on $[0,L_1]$}\label{conductance0L1}

We assume that the solution is nonzero over the support $[0,L_1]$ based
on our physical intuition for large $p$, where $L_1$ is yet unknown.
We use the general Tricomi solution with a single support in \eqref{formulatricomi}
with $a=0$, $b=L_1$ and $g(x)=p/2$ giving
\begin{align}\label{GeneraltricomiConductance}
\nonumber\varrho_p^\star(T) =-\frac{1}{\pi^2\sqrt{T(L_1-T)}} &\left[\frac{p}{2}~\mathrm{Pr}\int_0^{L_1}\frac{\sqrt{T^\prime
(L_1-T^\prime)}}{T-T^\prime}dT^\prime\right.\\
&\left. +B_1\right]
\end{align}
where $B_1$ is an arbitrary constant.
Evaluating the principal value integral
on the rhs of \eqref{GeneraltricomiConductance} we get
\begin{equation}\label{Solution0L1}
  \varrho_p^\star(T)=\frac{p}{2\pi\sqrt{T}}\sqrt{L_1-T}
\end{equation}
where the constant $B_1$ has been determined using the fact that the density
$\varrho_p^\star(T)$ must vanish at the upper edge $T=L_1$.
The normalization of $\varrho_p^\star(T)$ gives:
\begin{equation}\label{L1plarge}
L_1=\frac{4}{p},\qquad L_1<1\Rightarrow p>4
\end{equation}
As expected, this
solution holds for large values of $p$ (i.e. for a strong confining
potential).

Since the point $T=0$ belongs to the support $[0,L_1]$, we can put
$T=0$ in \eqref{StationaryPhase} to determine the constant $B$
\begin{equation}\label{Lagrange determined}
  B=2\int_0^1 \varrho_p^\star(T^\prime)\log T^\prime dT^\prime
\end{equation}
Substituting $B$ in \eqref{step2} gives the saddle point action
\begin{equation}\label{ActionConductanceSimplified}
S[\varrho_p^\star]=\frac{p}{2}\int_0^1 \varrho_p^\star(T)T
dT-\int_0^1\varrho_p^\star(T)\log(T)dT
\end{equation}
Performing the integrals using the explicit solution $\varrho_p^\star(T)$ 
\eqref{Solution0L1} gives a very simple expression, valid for $p>4$,
\begin{equation}\label{Action Spstar>4}
S[\varrho_{p}^\star]=3/2+\log p
\end{equation}

\begin{figure}[htb]
\begin{center}
\includegraphics[bb =-385   122   998   668,totalheight=0.14\textheight]{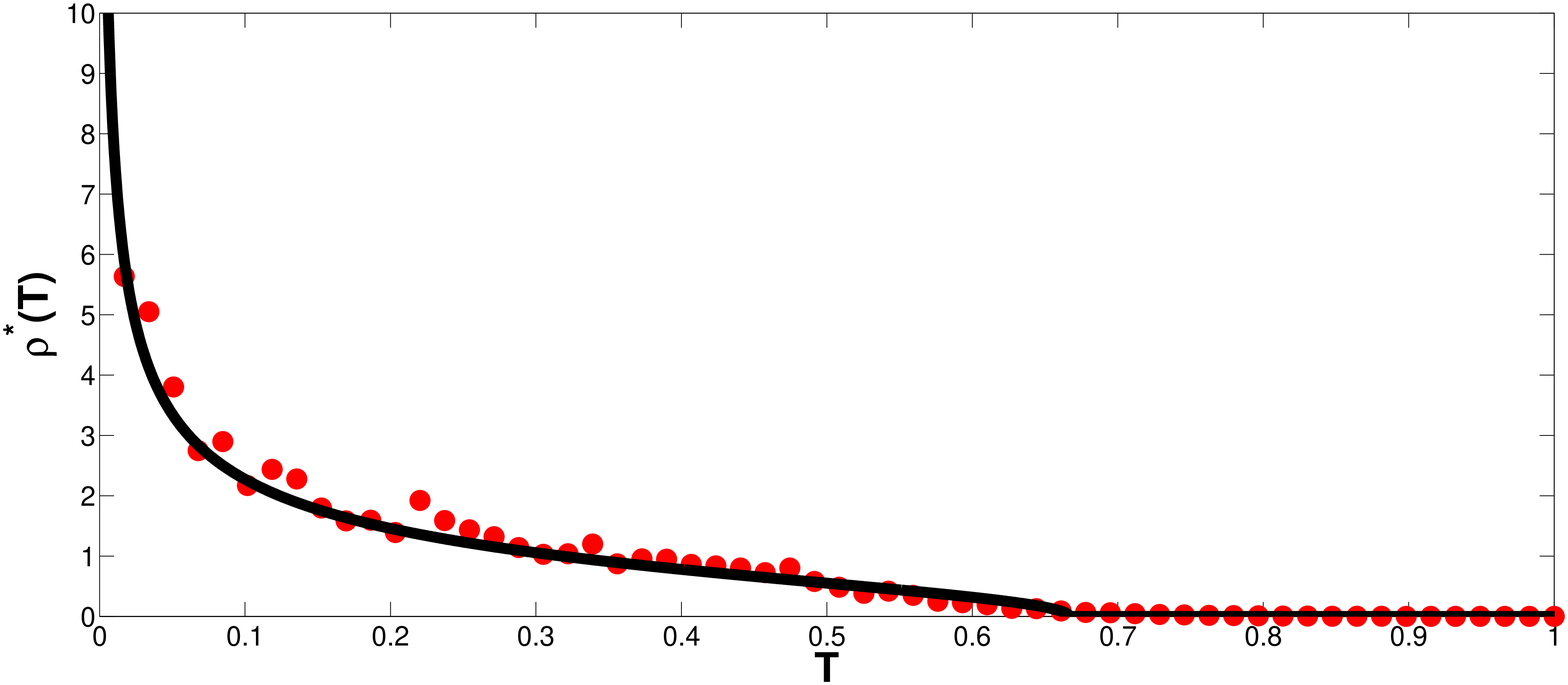}
\caption{(Color online). Density of transmission eigenvalues $T$ for
$N=5$ and $p=6$ (theory vs. numerics) for the conductance case.}\label{plargecondfig}
\end{center}
\end{figure}

In Fig. \ref{plargecondfig} we show the results from a Montecarlo simulation to
test the prediction \eqref{Solution0L1} for the average density of
eigenvalues in Laplace space for $p>4$.
The numerical density for $N$
eigenvalues (here and for all the subsequent cases) is obtained
as:
\begin{equation}\label{NumericalDensity}
  \varrho_p^\star(x_1)\simeq \frac{\Big\langle e^{-pN\sum_{i=1}^N
  x_i}\prod_{j<k}|x_j-x_k|^2\Big\rangle_{N-1}}{\Big\langle e^{-pN\sum_{i=1}^N
  x_i}\prod_{j<k}|x_j-x_k|^2\Big\rangle_N}
\end{equation}
where the average $\Big\langle\cdot\Big\rangle$ is taken over
$N-1$ random numbers ${x_2,\ldots,x_N}$ (numerator) with a flat
measure over $[0,1]$, with $x_1$ spanning the interval $(0,1)$. In
the denominator, the normalization constant is obtained with the
same procedure, this time averaging over $N$ random variables
uniformly drawn from $(0,1)$. In all cases, the agreement with the
theoretical results is fairly good already for $N=5$.

\subsection{Intermediate $p$: support on the full range $[0,1]$}
In this case, the solution of \eqref{Tricomisol} from
\eqref{formulatricomi} reads:
\begin{equation}\label{Solution01}
\varrho_p^\star(T)=\frac{p}{2\pi\sqrt{T(1-T)}}\left[K-T\right]
\end{equation}
The normalization of $\varrho_p^\star(T)$ determines $K=(4+p)/{2p}$.
Now, depending on whether $p>0$ or $p<0$, there are 2 positivity constraints
($\varrho_p^\star(T)>0$ everywhere) to take into account:
\begin{enumerate}
  \item if $p>0$, the positivity constraint $K-1>0$ at the upper edge $T=1$ implies
$p<4$.
  \item if $p<0$, the positivity constraint $K<0$ at the lower edge $T=0$ implies
$p>-4$.
\end{enumerate}
Thus the solution \eqref{Solution01} with support over the full alllowed range $[0,1]$
is valid for all $-4\le p\le 4$. 

Substituting this solution into the simplified action
\eqref{ActionConductanceSimplified} (which holds in this case as
well) gives:
\begin{equation}\label{Action Spstar<4}
  S[\varrho_p^\star]=-\frac{p^2}{32}+\frac{p}{2}+2\log 2
\end{equation}
Note that since this range $-4\le p\le 4$ includes, in particular the $p=0$ case,
we can use the expression in \eqref{Action Spstar<4} to evaluate the value
of the action at $p=0$ that will be required later in evaluating the large
deviation function via \eqref{AsymptoticDecay}. Putting $p=0$ in \eqref{Action Spstar<4}
gives
\begin{equation}
\label{action0}
 S[\varrho_0^\star]= 2 \log 2.
\end{equation}
 
\begin{figure}[htb]
\begin{center}
\includegraphics[bb =-385   122   998   668,totalheight=0.14\textheight]{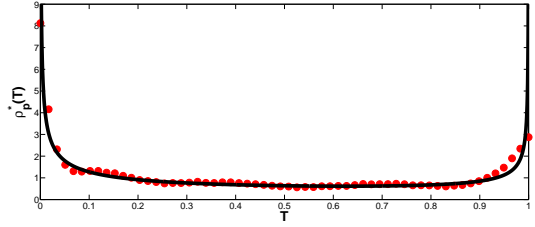}
\caption{(Color online). Density of transmission eigenvalues $T$ for
$N=5$ and $p=1$ (theory vs. numerics) for the conductance case.\label{P1figg}}
\end{center}
\end{figure}

In fig. \ref{P1figg}, we plot the analytical result for the density together with Montecarlo simulations for $N=5$ and $p=1$. 
 
\subsection{Large negative $p$: support on $[L_2,1]$}\label{caseC}
\begin{figure}[htb]
\begin{center}
\includegraphics[bb = -385   122   998   668,totalheight=0.14\textheight]{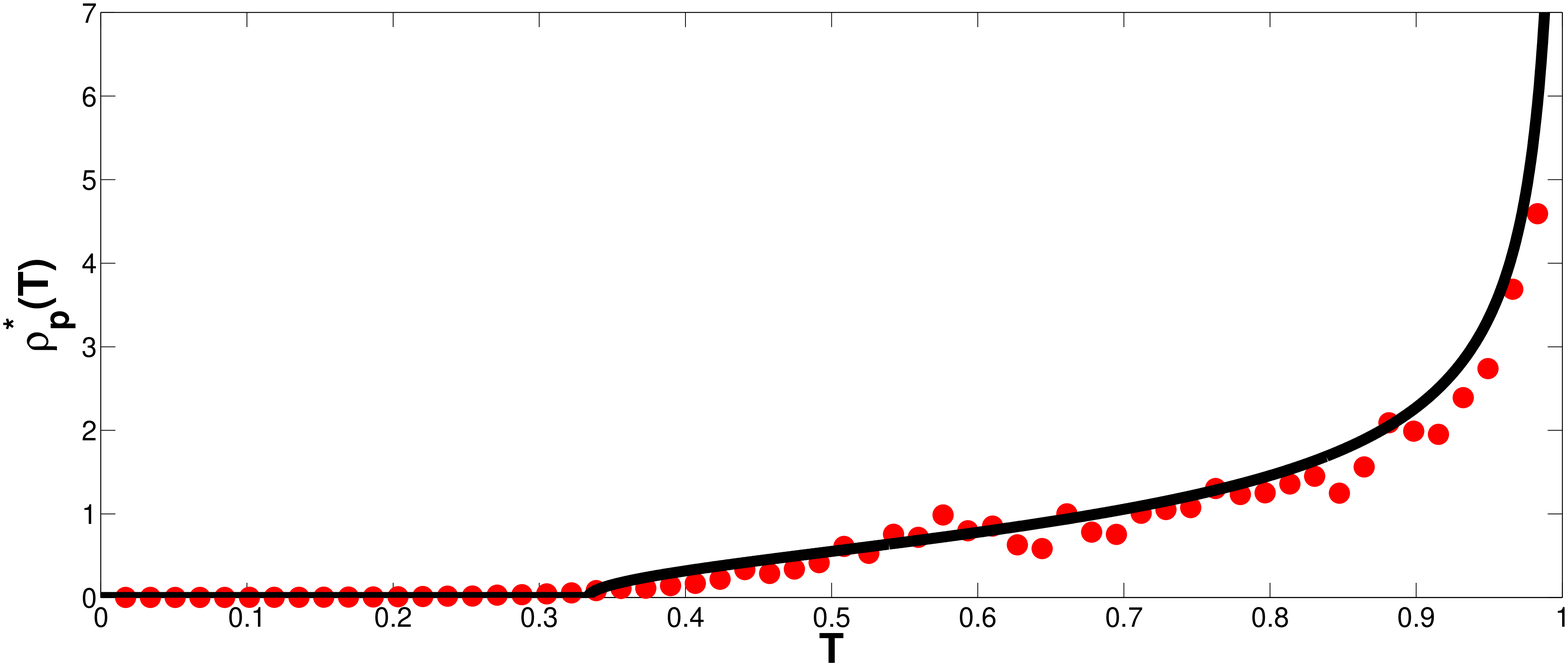}
\caption{(Color online). Density of transmission eigenvalues $T$ for
$N=5$ and $p=-6$ (theory vs. numerics) for the conductance case.}\label{Pmeno6}
\end{center}
\end{figure}
In this case, the solution of
\eqref{Tricomisol} reads:
\begin{equation}\label{SolutionL21}
\varrho_p^\star(T) = \frac{|p|}{2\pi\sqrt{1-T}} \sqrt{T-L_2},\qquad L_2=1-\frac{4}{|p|}
\end{equation}
The implications for the $p$-range are as follows:
\begin{equation}\label{implicationforp}
L_2>0\Rightarrow p<-4
\end{equation}

Note that we can no longer use the expression for the constant $B$ in \eqref{Lagrange determined}
since now the allowed range of the solution $[L_2,1]$ does not include the point $T=0$.
Instead, putting $T=1$ in \eqref{StationaryPhase}, we determine the value of $B$ as
\begin{equation}\label{Bnegative}
B=2\int_0^1\varrho_p^\star(x^\prime)\log|1-x^\prime|dx^\prime -p
\end{equation}
Substituting $B$ in \eqref{step2} we get a new expression for the action at
the saddle point
\begin{equation}\label{acz}
S[\varrho_p^\star]=\frac{p}{2}\int_0^1 dT T\, \varrho_p^\star(T)-\int_0^1 dT 
\varrho_p^\star(T)\,\log|1-T|+\frac{p}{2}
\end{equation}
Evaluating \eqref{acz} using the solution in \eqref{SolutionL21} gives the
saddle point action for $p<-4$
\begin{equation}\label{Action Spstar<-4}
  S[\varrho_p^\star]=3/2+ p + \log(-p)
\end{equation}

In fig. \ref{Pmeno6}, we plot the analytical result for the density together with Montecarlo simulations for $N=5$ and $p=-6$. 
\begin{figure}[htb]
\begin{center}
\includegraphics[bb =66   204   551   583,totalheight=0.25\textheight]{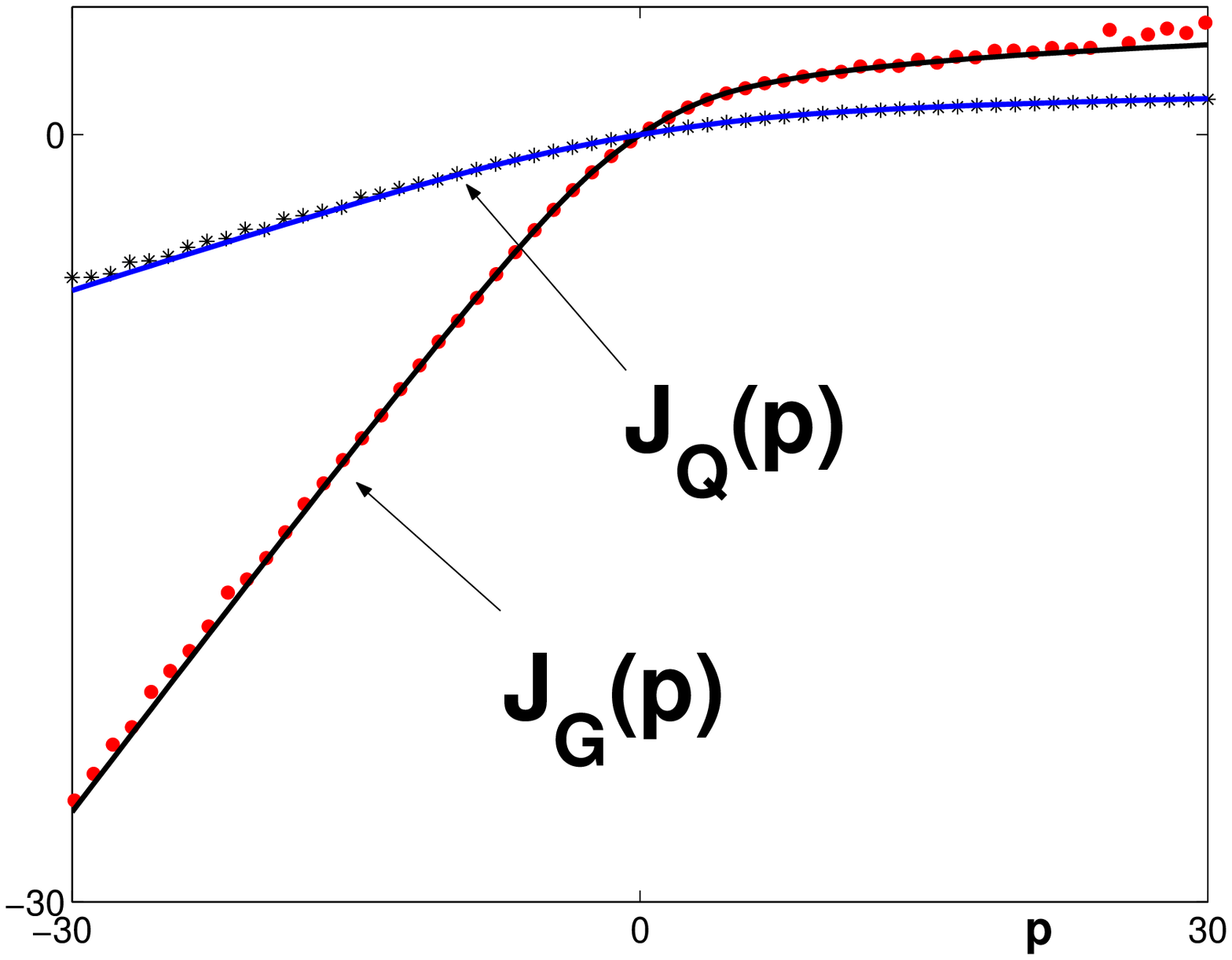}
\includegraphics[bb = 0 0 240 172,totalheight=0.30\textheight]{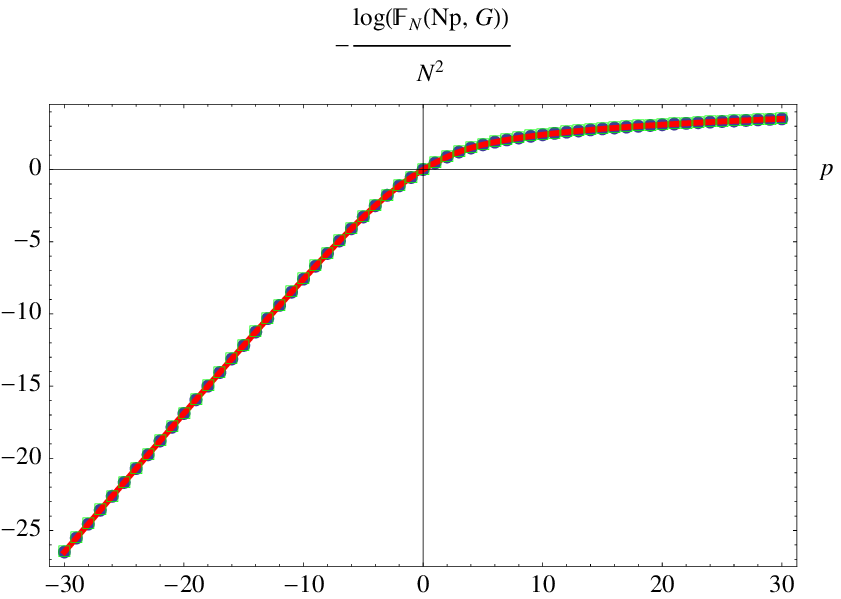}
\caption{(Color online). Left: $J_G(p)$ and $J_Q(p)$ vs. Montecarlo simulations (see eq. \eqref{NumericalJGdiP} and \eqref{AsymptoticDecayQ}). Right: 
our asymptotic predictions $J_G(p)$ (solid red line) is compared with the exact finite $N$ expression \eqref{detOK} from \cite{Kanz} for $N=4$ (green squares) and $N=10$ (blue dots). Already for $N=4$,
our large $N$ formula again matches the exact finite $N$ result with accuracy up to the second decimal digit over the full real $p$ range.\label{JgdiPfig}}
\end{center}
\end{figure}
\subsection{Comparison
with other theories and numerical simulations}\label{comp:sect}
As summarized in the next section, we obtained for the rate function of the conductance for $\beta=2$, defined as (see \eqref{limitrateGE}):
\begin{equation}
\Psi_G(x)=-\lim_{N\to\infty} \frac{\log \mathcal{P}_G(Nx,N)}{N^2}
\end{equation}
the following expression (limited to $x\in[1/2,1]$, given the symmetry $\Psi_G(x)=\Psi_G(1-x)$):
\begin{equation}\label{PsiGcomp}
   \Psi_G(x)=
  \begin{cases}
    8\left(x-\frac{1}{2}\right)^2 & 1/2\leq x\leq 3/4\\
    \frac{1}{2}-\log[4(1-x)] & 3/4\leq x< 1
  \end{cases}
\end{equation}
In \cite{Kanz}, Osipov and Kanzieper (OK) claim a different limiting law, namely:
\begin{equation}
\Psi_G^{\rm {OK}}(x) = 
\begin{cases}
 8\left(x-\frac{1}{2}\right)^2 &
1/2\le x\le 3/4 \\
4x-\frac{5}{2} & 3/4\le x< 1
\label{OKreg2}
\end{cases}
\end{equation}
and $\Psi_G^{\rm {OK}}(x)$ would approach the form in \eqref{PsiGcomp} (second line) only at the extreme
edge $x\to 1^{-}$ (over a narrow region of order $1/N$).

Which law is then correct? There is a conclusive way
to settle this dispute, namely to compare the two theoretical results
to a direct numerical simulation of $\mathcal{P}_G(G,N)$. 
We will present simulation results in Laplace
space which agree very well with our result on the Laplace
transform (see fig. \ref{JgdiPfig}, left panel, and equations \eqref{JGdiP} and \eqref{NumericalJGdiP} in next section) \emph{over the full range of real $p$ values},
as well as a convincing comparison with the \emph{exact} finite $N$ result for the same observable using the Hankel determinant representation \eqref{detOK} from \cite{Kanz} (see fig. \ref{JgdiPfig}, right panel). 
We shall argue below that OK asymptotic theory is instead unable to reproduce the tails of $J_G(p)$ for $|p|>4$, which are responsible for long power-law tails in the rate function $\Psi_G(x)$.
Since, however, working in the Laplace space may not appear conclusive as far as the real-space rate function $\Psi_G(x)$ is concerned,
it would be better if
one could perform a simulation directly for $\mathcal{P}_G(G,N)$ and not just for its
Laplace transform.

Indeed, it turns out to be quite easy to simulate directly $\mathcal{P}_G(G,N)$
using an elementary and standard Monte Carlo Metropolis algorithm
which we describe below.
\vskip 0.4cm

\noindent{\bf Monte Carlo method:}
\vskip 0.2cm

The main problem is to compute the distribution of the conductance $G$ 
which, for a fixed number of channels $N$ and $\beta=2$, is given
by the multiple integral 
\begin{equation}
\mathcal{P}_G(G,N)\!= \!\! A_N\, \!\!\! \int_0^1 \prod_{i=1}^N dT_i\prod_{i<j}(T_i-T_j)^2\, 
\delta\left(\sum_{i=1}^N T_i-G\right)
\label{mult1}
\end{equation}
where the prefactor $A_N$ is set by the normalization: $\int_0^{\infty} 
\mathcal{P}_G(G,N)dG=1$ and is known exactly for all $N$ (see eq. \eqref{A_N}).

\begin{figure}[htb]
\begin{center}
\includegraphics[bb =49 54 511 499,totalheight=0.30\textheight]{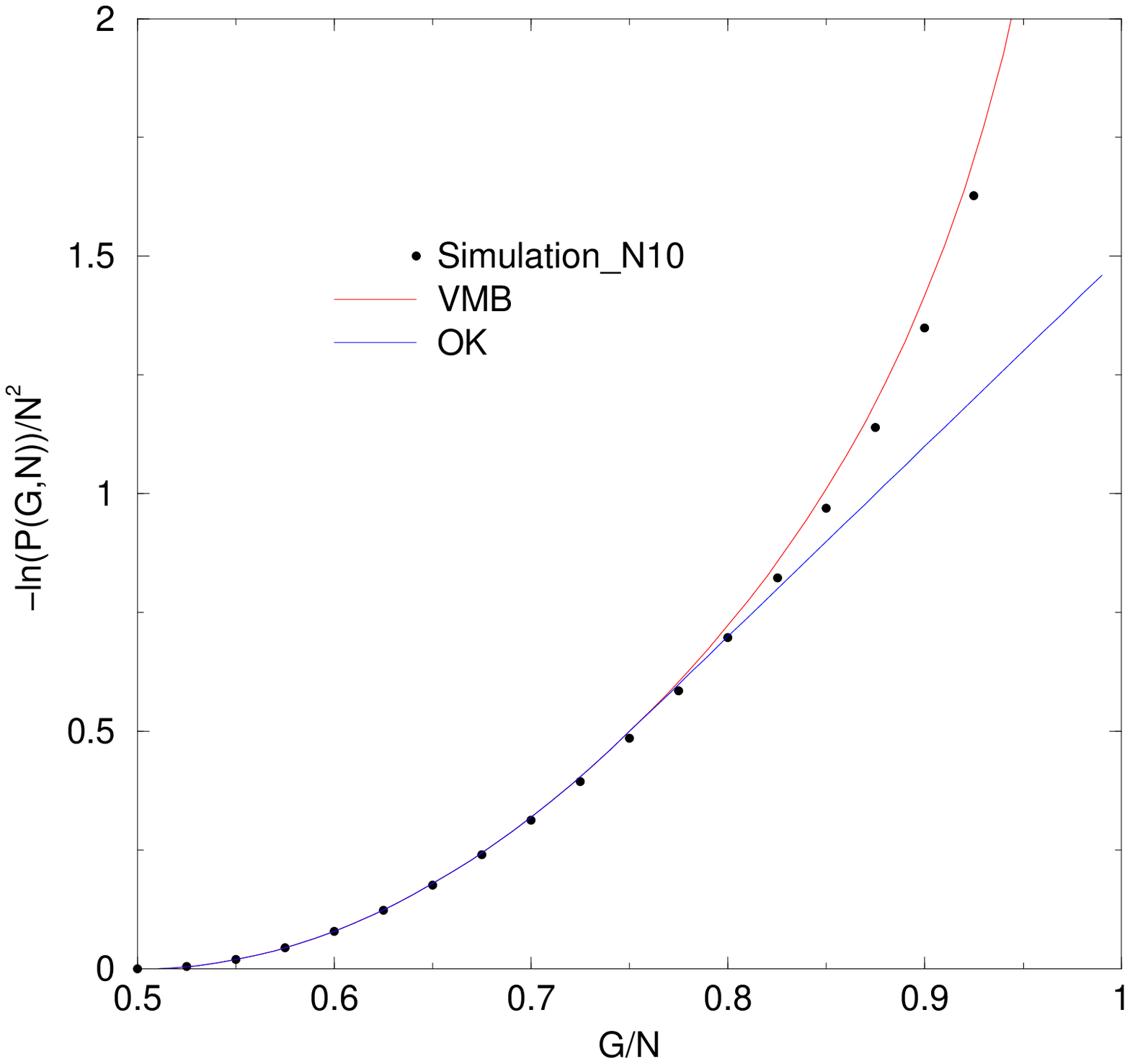}
\includegraphics[bb =49 53 511 499,totalheight=0.30\textheight]{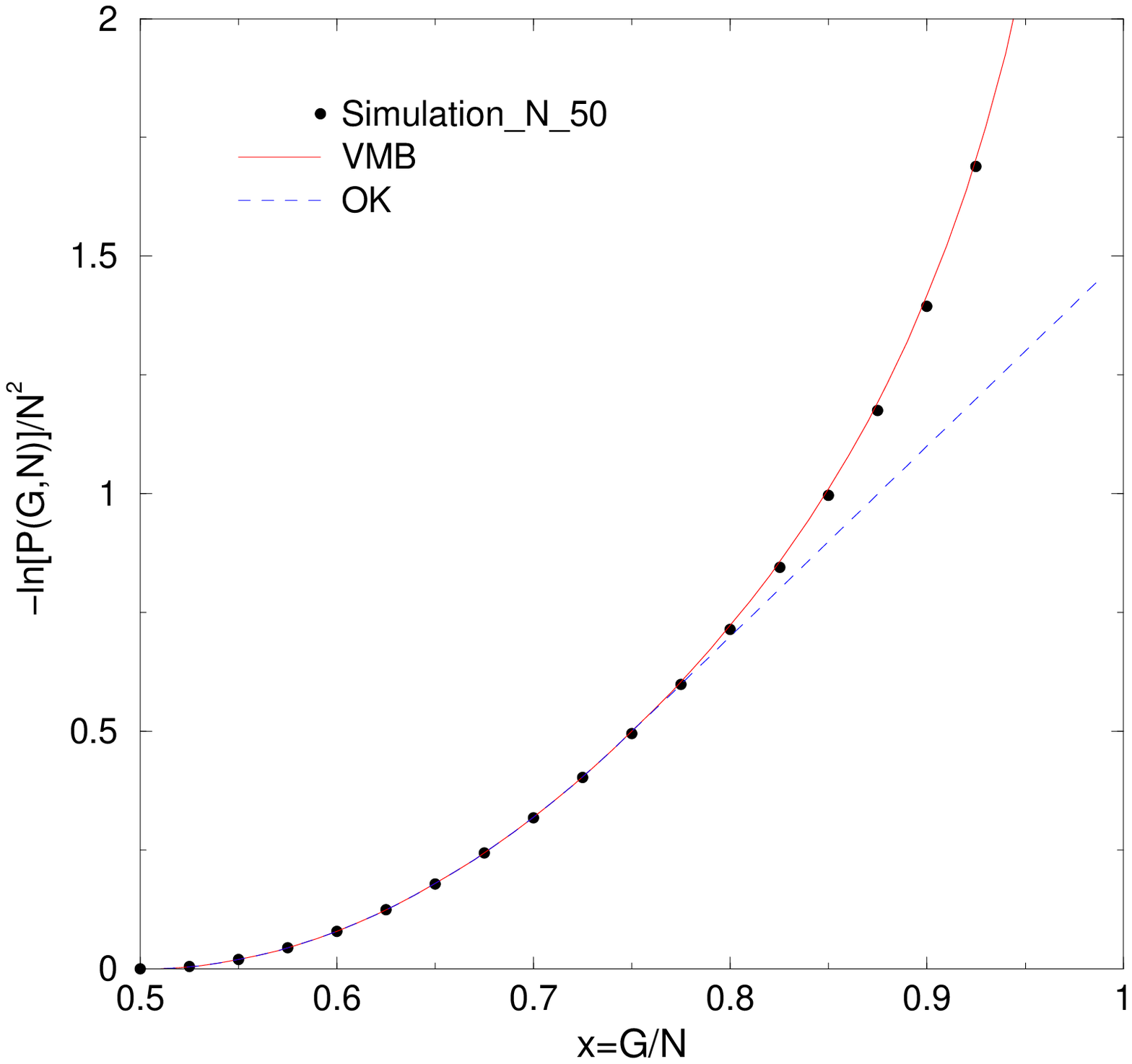}
\caption{(Color online). Comparison between Montecarlo simulations for the rate function $\Psi(x)=-\lim_{N\to\infty}\log \mathcal{P}_G(Nx,N)/N^2$ as a function of $x=G/N$ for $\beta=2$ (black dots), our asymptotic theory (VMB)
(continuous red line) and the theory by Osipov-Kanzieper (OK) (dashed blue line). Simulations are performed for $N=10$ (left) and $N=50$ (right), which allows to appreciate the convergence to our
theoretical curve as $N$ is increased.\label{RateComp}}
\end{center}
\end{figure}

To employ the Monte Carlo method, we first write the integrand (the 
Vandermonde term)
in \eqref{mult1}
\begin{equation}
\prod_{i<j} (T_i-T_j)^2= e^{-E\{T_i\}};\hspace{4pt}
E\{T_i\}= -\sum_{i\ne j} \ln\left(|T_i-T_j|\right).
\label{coulomb1}
\end{equation}
Thus, one can interpret $0\le T_i\le 1$ as the position of an $i$-th charge
in a one dimensional box $[0,1]$ and the charges interact via the logarithmic 
Coloumb energy
$-E\{T_i\}$. This Coulomb gas is
in thermal equilibrium with a Gibbs weight $\exp[-E\{T_i\}]$
for any configuration $\{T_i\}$, where the inverse temperature
is set to $1$. 

It is then very easy and standard to simulate the equilibrium
properties of this gas via a Monte Carlo method~\cite{Krauth}. We start
from any configuration $\{T_i\}$. We pick a particle, say the $i$-th one, at 
random
and attempt to move its position by an amount $\delta$: $T_i\to T_i+\delta$.
This move causes a change in energy $\Delta E$ of the gas.
According to the standard Metropolis algorithm~\cite{Krauth}, the move is 
accepted with
probability $e^{-\Delta E}$ if $\Delta E>0$ and with probability $1$
if $\Delta E<0$. The move is rejected if the new position
$T_i+\delta$ is outside the box $[0,1]$. This Metropolis move
guarantees that after a large number of microscopic moves,
the system reaches the stationary distribution with the
correct Boltzmann weight $e^{-E\{T_i\}}$.

We wait for a long enough time to ensure that the system has indeed
reached equilibrium. After that we let the system evolve
according to these microscopic moves and construct the normalized histogram
$\mathcal{P}_G(G,N)$ of $G= \sum_{i=1}^N T_i$. Once again, we are guaranteed
that $T_i$'s are sampled with the correct equilibrium weight.
This procedure allows us to simulate fairly large systems. To obtain good statistics for the distribution over the full
range of $x=G/N$, i.e., over $1/2\le x\le 1$, we implement
an iterative conditional sampling method,
used in other contexts before \cite{verg2}, that allows us
to generate events with extremely small probabilities
at the far tail of the distribution~\cite{CelineFoot}.

In Fig. \ref{RateComp},
we plot $\Psi(x)=-\ln[\mathcal{P}_G(G,N)]/N^2$ vs the scaling variable $x=G/N$
for $1/2\le x\le 1$ for $N=50$. The black dots show the
simulation points. The red (solid) line shows our (VMB) result
in \eqref{PsiGcomp}, while the 
blue (dashed) line shows the OK prediction \eqref{OKreg2}.
Clearly
the two results agree with each other, as well as with the simulations,
in the Gaussian regime $1/2\le x\le 3/4$. However, in the outer
region $3/4\le x< 1$, while the VMB result in \eqref{PsiGcomp}
is in perfect agreement with the simulation results, the OK
result in \eqref{OKreg2} deviates widely from them.
{\em This proves conclusively that
in the regime $3/4\le x<  1$, the VMB result \eqref{PsiGcomp}
is correct and the OK result \eqref{OKreg2} is incorrect.}

There is another way to see that the OK asymptotics
can not be correct. Since OK theory stems from the asymptotic analysis of the 
following integral representation for the probability distribution of conductance (see eq. 22 of \cite{Kanz}):
\begin{equation}\label{kanz1}
\mathcal{P}_G^{\rm{(OK)}}(G,N)=\frac{2 N^{1/4}}{\Gamma(1/8)}\sqrt{\frac{2}{\pi}}\int_0^\infty d\lambda\frac{e^{-N^2\left(\lambda+\frac{2 \eta^2}{1+2\lambda}\right)}}{\lambda^{7/8}\sqrt{1+2\lambda}}
\end{equation}
where $\eta = 2 (G/N)-1$, we can now compute from \eqref{kanz1} the \emph{same} observable we considered here, namely:
\begin{equation}
J_G(p)=-\lim_{N\to\infty}\frac{\log\mathbf{F}_N(Np;G)}{N^2}
\end{equation}
for $p\in\mathbb{R}$ and $p\sim\mathcal{O}(1)$ for $N\to\infty$ (see equations \eqref{Laplacez} and \eqref{limitJP}), and thus compare once again the large $N$ predictions of VMB and OK theories against i) numerical simulations, and ii) the exact finite $N$ result, which is fortunately available (see fig. \ref{JgdiPfig}).

Inserting \eqref{kanz1} into the definition of the Laplace transform \eqref{Laplacez}, we obtain after simple algebraic steps:
\begin{align}\label{eventuallyLaplace}
\nonumber\mathbf{F}_N(Np;G) &=\frac{N^{1/4}}{\Gamma(1/8)}\exp\left(-N^2\left(\frac{p}{2}-\frac{p^2}{32}\right)\right)\times\\
&\times\underbrace{\int_0^\infty d\lambda
\frac{e^{-N^2\lambda\left(1-p^2/16\right)}}{\lambda^{7/8}}}_{\phi_N(p)}
\end{align}
The integral $\phi_N(p)$ clearly does not converge for $|p|\geq 4$, with the consequence that OK integral representation \eqref{kanz1} fails
to reproduce the tails of both: 
\begin{itemize}
\item Montecarlo simulations of $J_G(p)$ (see fig. \ref{JgdiPfig}, left panel), 
\item the exact finite $N$ result \cite{Kanz} for the Laplace transform in terms of a Hankel determinant (see fig. \ref{JgdiPfig}, right panel):
\begin{equation}\label{detOK}
\mathbf{F}_N(Np;G)=\frac{N!}{c_N}\det\left[(-\partial _z)^{j+k} \frac{1-e^{-z}}{z}\Big|_{z=Np}\right]_{0\leq j,k\leq N-1},\qquad c_N=\prod_{j=0}^{N-1}\frac{\Gamma(j+2)\Gamma^2(j+1)}{\Gamma(j+N+1)}
\end{equation}
\end{itemize}
which evidently \emph{do} exist and are perfectly captured instead by our approach in both cases.

\emph{Within the range of validity $|p|<4$}, the integral $\phi_N(p)$ can be evaluated and gives eventually:
 \begin{equation}\label{OKresult}
\mathbf{F}_N(Np;G)=\frac{\exp\left(-N^2 (p/2-p^2/32)\right)}{(1-p^2/16)^{1/8}},\qquad\qquad\mbox{if   } |p|<4
\end{equation}
 Note that \emph{all} $N$-dependence has completely dropped out from the prefactor, leaving us with two perfectly compatible (even though apparently discordant at first glance) consequences:
 \begin{enumerate}
 \item According to OK theory, the limit
 \begin{equation}
J_G(p)=-\lim_{N\to\infty}\frac{\log\mathbf{F}_N(Np;G)}{N^2}=p/2-p^2/32\qquad\mbox{for  }|p|<4
\end{equation}
which is correct but incomplete (as their integral has nothing to say about the tails $|p|\geq 4$).
\item Not surprisingly then, from the OK theory, the leading $(1/N)$ order of the cumulants of the distribution, computed through the formula \eqref{cum1}:
\begin{equation}\label{cumulantsnew}
\kappa_\ell(G)=\left(\frac{-1}{N}\right)^\ell\frac{\partial^\ell}{\partial p^\ell}\log \mathbf{F}_N(Np;G)\Big|_{p=0}
\end{equation} 
is correctly reproduced (see also \cite{savinnew} for an independent calculation of such cumulants, which is in perfect agreement with OK result).
 \end{enumerate}
 In summary, the OK integral representation in \eqref{kanz1} is only adequate around the Gaussian peak (the $p=0$ neighborhood in Laplace space, \emph{which is exactly the only region
 which cumulants probe (see \eqref{cumulantsnew})}), and in this neighborhood it has the merit of producing the correct leading $1/N$ term of the expansion of such cumulants (unattainable by
 our method) and confirmed independently in \cite{savinnew}. Outside this region, however, the OK integral representation is invalid and the asymptotic analysis of it \emph{outside its range of validity} obviously produces an incorrect result.   
 Note also that, for any fixed $N$ (however large), one has from \eqref{kanz1} that $\mathcal{P}_G^{\rm{(OK)}}(G=0,N)=\mathcal{P}_G^{\rm{(OK)}}(G=N,N)>0$ strictly, while it is well-known
 that the density must vanish identically at the edges \cite{sommers,savinnew,mello22} \emph{for any $N$}. This simple observation rules out the claims of exactness of \eqref{kanz1} in \cite{Kanz}.

\subsection{Final results for the conductance case}

To summarize, the density of eigenvalues (solution
of the saddle point equation \eqref{Tricomisol}) has the following
form:
\begin{widetext}
\begin{equation}\label{DensitySummary}
  \varrho_p^\star(T)=
  \begin{cases}
  \frac{p}{2\pi\sqrt{T(1-T)}}\left[\frac{4+p}{2p}-T\right] &
  \qquad 0\leq T\leq 1 \qquad -4\leq p\leq 4\\
\frac{p}{2\pi\sqrt{T}}\sqrt{\frac{4}{p}-T} &
  \qquad 0\leq T\leq 4/p \qquad p\geq 4\\
  \frac{|p|}{2\pi\sqrt{1-T}} \sqrt{T-(1-4/|p|)} &
  \qquad 1-4/|p|\leq T\leq 1 \qquad p\leq -4
  \end{cases}
\end{equation}
\end{widetext}
One may easily check that $\varrho_p^\star(T)$ is continuous at $p=\pm 4$,
but develops two phase transitions characterized by different
supports.

The action at the saddle point is given by:
\begin{equation}\label{Spstar summary}
  S[\varrho_p^\star]=
  \begin{cases}
  -\frac{p^2}{32}+\frac{p}{2}+2\log
  2 &
 -4\leq p\leq 4\\
3/2+\log p &
   p\geq 4\\
  3/2+p + \log(-p) &
  p\leq -4
  \end{cases}
\end{equation}
which is again continuous at $p=\pm 4$.

Using the above expressions for the saddle point action
and the result in \eqref{action0}, the expression for the 
free energy difference $J_G(p)=S[\varrho_p^\star]-S[\varrho_0^\star]$ follows from \eqref{AsymptoticDecay}:
\begin{equation}\label{JGdiP}
  J_G(p)=
  \begin{cases}
-\frac{p^2}{32}+\frac{p}{2} &
 -4\leq p\leq 4\\
3/2+\log (p/4) &
   p\geq 4\\
  3/2+p + \log(-p/4) &
  p\leq -4
  \end{cases}
\end{equation}

Using this expression for $J_G(p)$ in the Legendre transform \eqref{legendre1} and
maximizing gives
the exact expression for the rate function
\begin{equation}\label{PsiG}
   \Psi_G(x)=
  \begin{cases}
    8\left(x-\frac{1}{2}\right)^2 & \frac{1}{4}\leq x\leq \frac{3}{4} \\
    \frac{1}{2}-\log(4x) & 0\leq x\leq \frac{1}{4}\\
    \frac{1}{2}-\log[4(1-x)] & \frac{3}{4}\leq x\leq 1
  \end{cases}
\end{equation}
From this formula, one can derive the leading behavior of the
tails of $\mathcal{P}_G(G,N)$ as:
\begin{align}\label{LeadingAsymptotics}
  \mathcal{P}_G(G,N) &\stackrel{G\to
  0}{\approx}\exp\left\{-\frac{\beta}{2}N^2\left[-\log\left(4G/N\right)\right]\right\}=
  G^{\beta N^2/2}\\
  \nonumber\mathcal{P}_G(G,N) &\stackrel{G\to
  N}{\approx}\exp\left\{-\frac{\beta}{2}N^2\left[-\log\left(4(1-G/N)\right)\right]\right\}=\\
\nonumber &= (N-G)^{\beta N^2/2}\\
\end{align}
in agreement with \cite{sommers,savinnew,mello22}.

The most interesting feature of \eqref{PsiG} is the appearance of
discontinuities in higher-order derivatives at the critical
points: more precisely, the third derivative of $\Psi_G(x)$ is
discontinuous at $x=1/4$ and $x=3/4$. In fact, Sommers et al \cite{sommers} found 
that for finite $N_1,N_2$ there are several non-analytical points.
Only two of them survive
to the leading order $N\to\infty$ and, in our picture, these
correspond to a physical phase transition in Laplace space.

In summary, the following exact limit holds:
\begin{equation}\label{LimitLargeDeviations}
  \lim_{N\to\infty}\left[-\frac{2\log\mathcal{P}_G(Nx,N)}{\beta
  N^2}\right]=\Psi_G(x)
\end{equation}
where the rate function $\Psi_G(x)$ is given in \eqref{PsiG}.

The free energy difference in Laplace space \eqref{JGdiP} for large $N$ has been compared
with:
\begin{enumerate} 
\item Montecarlo simulations over the range $p=(-30,30)$, which already
for $N=5$ show an excellent agreement (see fig. \ref{JgdiPfig}, left panel). For a given $p$ between
$-30$ and $30$, the numerical $J_G(p)$ (and analogously for $J_Q(p)$) is computed as:
\begin{equation}\label{NumericalJGdiP}
  J_G(p)\simeq \frac{\Big\langle e^{-pN\sum_{i=1}^N
  x_i}\prod_{j<k}|x_j-x_k|^2\Big\rangle_{N}}{\Big\langle \prod_{j<k}|x_j-x_k|^2\Big\rangle_N}
\end{equation}
where the average is taken over $N$ random variables $\{x_i\}$
drawn from a uniform distribution over $(0,1)$.
\item the \emph{exact} finite $N$ result from \cite{Kanz} for the Laplace transform of the density in terms of a Hankel determinant (see \eqref{detOK}). In fig. \ref{JgdiPfig} (right panel),
we plot our asymptotic result $J_G(p)$ \eqref{JGdiP} together with $-\log\mathbf{F}_N(Np;G)/N^2$, where $\mathbf{F}_N(Np;G)$ is the exact finite $N$ result \eqref{detOK} for the Laplace transform from \cite{Kanz},
for $-30\leq p\leq 30$ and $N=4,10$. Already for $N=4$, our $J_G(p)$ reproduces the exact finite $N$ formula with an accuracy of two decimal digits.
\end{enumerate}
\section{Distribution of the shot noise}

The dimensionless shot noise is defined as $P=\sum_{i=1}^N
T_i(1-T_i)$ \cite{lesovik,ya}. It is convenient to rewrite it in
the form $P= N/4-\sum_{i=1}^N (1/2-T_i)^2=N/4-Q$. The
probability distributions of $P$ and $Q$ are related by:
\begin{equation}\label{ProbabilityFandQ}
  \mathcal{P}_P(P,N)=\mathcal{P}_Q\left(\frac{N}{4}-P,N\right)
\end{equation}
\begin{figure}[htb]
\begin{center}
\includegraphics[bb = 14 14 527 399,totalheight=0.27\textheight]{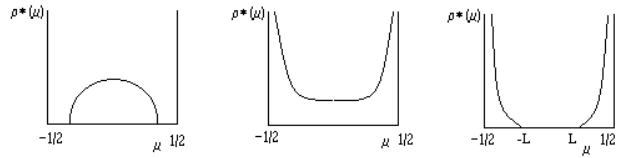}
\caption{Density of the auxiliary $Q$ (schematic).\label{sketchfano}}
\end{center}
\end{figure}
It is also necessary to make the change of variable in the joint pdf
\eqref{jpd transmission First} $\mu_i=1/2-T_i$, so that
$-1/2\leq\mu_i\leq 1/2$. The joint pdf \eqref{jpd transmission First}
expressed in terms of the new variables $\mu_i$ reads:
\begin{equation}\label{jpdmu}
 P(\mu_1,\ldots,\mu_N)=A_N\prod_{j<k}|\mu_j-\mu_k|^\beta
\prod_{i=1}^N\left(\frac{1}{2}-\mu_i\right)^{\frac{\beta}{2}-1}
\end{equation}
and we are interested in the large $N$ decay of the logarithm of
$\mathcal{P}_Q(Q,N)$, where $Q=\sum_i \mu_i^2$. We have:
\begin{widetext}
\begin{equation}\label{DistributionQ}
\mathcal{P}_Q(Q,N)=A_N\int_{-1/2}^{1/2}\cdots\int_{-1/2}^{1/2}
d\mu_1\cdots d\mu_N \exp\left(\frac{\beta}{2}\sum_{j\neq
k}\log|\mu_j-\mu_k|+\left(\frac{\beta}{2}-1\right)\sum_{i=1}^N
\log \left(\frac{1}{2}-\mu_i\right)\right)\delta\left(\sum_{i=1}^N
\mu_i^2-Q\right)
\end{equation}
\end{widetext}
Again, taking the Laplace transform and converting multiple
integrals to functional integrals we obtain:
\begin{equation}\label{MultipleToFunctionalQ}
 \int_0^\infty \mathcal{P}_Q(Q,N)e^{-\frac{\beta}{2}NpQ}dQ=
  A_N\int\mathcal{D}[\varrho]e^{-\frac{\beta}{2}N^2 S[\varrho_p]}
\end{equation}
where for notational simplicity we keep the same symbols $\varrho_p$
and $S$ as before. Of course, the new action $S$ reads:
\begin{widetext}
\begin{equation}\label{S_pQ}
  S[\varrho_p]=p\int_{-1/2}^{1/2} \varrho_p(\mu)\mu^2 d\mu-\int_{-1/2}^{1/2}\int_{-1/2}^{1/2} d\mu d\mu^\prime
  \varrho_p(\mu)\varrho_p(\mu^\prime)\log|\mu-\mu^\prime|+C\left[\int_{-1/2}^{1/2} \varrho_p(\mu)d\mu-1\right]
\end{equation}
\end{widetext}
where $C$ is the new Lagrange multiplier enforcing the normalization of the charge density
to unity.

The stationary point of the action $S$ is determined by:
\begin{equation}
\frac{\delta S[\varrho_p]}{\delta\varrho_p}=0
\end{equation}
yielding:
\begin{equation}\label{Ceq}
p {\mu}^2+C=2\int_{-1/2}^{1/2}d{\mu}^\prime\varrho_p^\star({\mu}^\prime)\log|\mu-{\mu}^\prime|
\end{equation}
Taking one more derivative with respect to $\mu$, we get to
the following Tricomi equation:
\begin{equation}\label{TricomiQ}
  p\mu=\mathrm{Pr}\int_{-1/2}^{1/2}\frac{\varrho_p^\star({\mu}^\prime)}{\mu-{\mu}^\prime}d{\mu}^\prime
\end{equation}

In terms of the solution $\varrho_p^\star(\mu)$ of \eqref{TricomiQ},
the action \eqref{S_pQ} can be simplified as:
\begin{equation}\label{spQsimplified}
  S[\varrho_p^\star]=\frac{p}{2}\int_{-1/2}^{1/2} \varrho_p^\star(\mu){\mu}^2
  d\mu-\frac{C}{2}
\end{equation}
where the value of the constant $C$ is determined from \eqref{Ceq} by attributing a value to $\mu$ within the support of the solution.
As in the conductance case, we can write the asymptotic decay of
$Q$ as:
\begin{widetext}
\begin{equation}\label{AsymptoticDecayQ}
\int_0^\infty \mathcal{P}_Q(Q,N)e^{-\frac{\beta}{2}NpQ}dQ\approx
\exp\left(-\frac{\beta}{2}N^2\{S[\varrho_p^\star]-S[\varrho_0^\star]\}\right)=\exp\left(-\frac{\beta}{2}N^2
J_Q(p)\right)
\end{equation}
\end{widetext}
Again, in order to solve \eqref{TricomiQ} we need first to foresee
the structure of the allowed support for $\varrho_p^\star(\mu)$. This time, the
symmetry constraint $\varrho_p^\star(\mu)=\varrho_p^\star(-\mu)$ reduces
the possible behaviors of $\varrho_p^\star(\mu)$ to the following
three cases: I) $\varrho_p^\star(\mu)$ has compact support $[-L,L]$
with $L<1/2$, or II) $\varrho_p^\star(\mu)$ has non-compact support
$(-1/2,1/2)$, or III) the support of $\varrho_p^\star(\mu)$ is the
union of two disjoint semi-compact intervals $(-1/2,-L]\cup
[L,1/2)$ with $L>0$ (see Fig. \ref{sketchfano}). We analyze the three cases separately.

\subsection{Support on $[-L,L]$ with $L<1/2$}\label{Fano-LL}
\begin{figure}[htb]
\begin{center}
\includegraphics[bb = -385   122   998   668,totalheight=0.14\textheight]{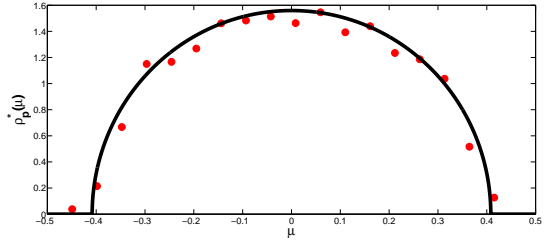}
\caption{(Color online). Density of shifted transmission eigenvalues $\mu$ for
$N=6$ and $p=12$ (theory vs. numerics) for the shot noise case.}
\label{fanoA}
\end{center}
\end{figure}
The general solution of \eqref{TricomiQ} in this case is given by:
\begin{equation}\label{rhostarFanowithconstants}
\varrho_p^\star(\mu)=\frac{p}{\pi\sqrt{L^2-{\mu}^2}}[c_1-{\mu}^2]
\end{equation}
The constant $c_1$ is clearly determined as $c_1=L^2$ by the condition
that $\varrho_p^\star(\pm L)=0$. Thus, the solution within the
bounds $[-L,L]$ with $L<1/2$ is the semicircle:
\begin{equation}\label{Semicircle}
\varrho_p^\star(\mu)=\frac{p}{\pi}\sqrt{L^2-{\mu}^2}
\end{equation}
where the edge point $L$ is determined by the normalization
condition $\int_{-L}^L \varrho_p^\star(\mu)d\mu=1$. This gives:
\begin{equation}
L=\sqrt{\frac{2}{p}},\qquad L<1/2\Rightarrow p>8
\end{equation}
So, eventually (see Fig. \ref{fanoA}):
\begin{equation}\label{SemicircleAfterConditions}
\varrho_{p}^\star(\mu)=\frac{p}{\pi}\sqrt{\frac{2}{p}-{\mu}^2},
\qquad -\sqrt{\frac{2}{p}}\leq \mu\leq\sqrt{\frac{2}{p}},\qquad p>8
\end{equation}
Evaluating the action \eqref{spQsimplified} (after determining the constant $C$ from \eqref{Ceq} by putting $\mu=0$ there) we get for $p>8$:
\begin{equation}\label{spQsimplifiedEvaluated p>8}
  S[\varrho_p^\star]=\frac{3}{4}+\frac{1}{2}\log
  2+\frac{1}{2}\log p
\end{equation}
From \eqref{AsymptoticDecayQ}, the value of
$J_Q(p)=S[\varrho_p^\star]-S[\varrho_0^\star]$ 
(still using $S[\varrho_0^\star]=2\log 2$) for $p\geq 8$
is given by:
\begin{equation}\label{JQp>8}
J_Q(p)=\frac{3}{4}+\frac{1}{2}\log\left(\frac{p}{8}\right)
\end{equation}
Again, the rate function $\Psi_Q(x)$ is given by the inverse
Legendre transform of \eqref{JQp>8}, i.e.:
\begin{equation}\label{RateFanoLeg}
 \Psi_Q(x)=\max_p\left[-xp+J_Q(p)\right]=\frac{1}{4}-2\log 2
 -\frac{1}{2}\log x
\end{equation}
valid for $0\leq x \leq 1/16$. From
\eqref{ProbabilityFandQ}, we have the following relation among the
rate functions for $Q$ and $P$:
\begin{equation}\label{RateQandP}
   \Psi_P(x)=\Psi_Q\left(\frac{1}{4}-x\right)
\end{equation}
implying for $\Psi_P(x)$ the following expression:
\begin{equation}\label{Psi_P}
  \Psi_P(x)=\frac{1}{4}-2\log
  2-\frac{1}{2}\log\left(\frac{1}{4}-x\right)\qquad\text{for}\quad
  \frac{3}{16}\leq x\leq \frac{1}{4}
\end{equation}

In fig. \ref{fanoA} we plot the theoretical density of shifted transmission eigenvalues together with Montecarlo simulations for $N=6$ and $p=12$.

\subsection{Support on $(-1/2,1/2)$}\label{Fano-1/2}
The general solution of \eqref{TricomiQ} in this case is given by:
\begin{equation}\label{rhostarFanowithconstants-1/2}
\varrho_p^\star(\mu)=\frac{p}{\pi\sqrt{1/4-{\mu}^2}}[b_1-{\mu}^2]
\end{equation}
The constant $b_1$ is determined by the normalization
condition $\int_{-1/2}^{1/2} \varrho_p^\star(\mu)d\mu=1$. This gives
$b_1=1/p+1/8$. In turn, the positivity constraint for the density
implies $-8<p<8$. We then get:
\begin{widetext}
\begin{equation}\label{SemicircleAfterConditions1}
\varrho_{p}^\star(\mu)=\frac{p}{\pi\sqrt{1/4-{\mu}^2}}
\left[\frac{1}{p}+\frac{1}{8}-{\mu}^2\right],\qquad -\frac{1}{2}< \mu<\frac{1}{2},\qquad -8< p< 8
\end{equation}
\end{widetext}

Evaluating the action \eqref{spQsimplified} gives for $-8< p< 8$:
\begin{equation}\label{spQsimplifiedEvaluated -8<p<8}
  S[\varrho_p^\star]=\frac{p}{8}-\frac{p^2}{256}+2\log 2
\end{equation}
From \eqref{AsymptoticDecayQ}, the value of
$J_Q(p)=S[\varrho_p^\star]-S[\varrho_0^\star]$ for $-8\leq
p\leq 8$ is given by:
\begin{equation}\label{JQp<8}
J_Q(p)=-\frac{p^2}{256}+\frac{p}{8}
\end{equation}
Again, the rate function $\Psi_Q(x)$ is given by the inverse
Legendre transform of \eqref{JQp<8}, i.e.:
\begin{equation}\label{RateFanoLeg<8}
 \Psi_Q(x)=\max_p\left[-xp+J_Q(p)\right]=64\left(x-\frac{1}{8}\right)^2
\end{equation}
valid for $1/16\leq x \leq 3/16$. From the relation
\eqref{RateQandP}, we have for $\Psi_P(x)$ the following
expression:
\begin{equation}\label{Psi_P<3/16}
  \Psi_P(x)=64\left(\frac{1}{8}-x\right)^2\qquad\text{for}\quad
  \frac{1}{16}\leq x\leq \frac{3}{16}
\end{equation}

\begin{figure}[htb]
\begin{center}
\includegraphics[bb =-385   122   998   668,totalheight=0.14\textheight]{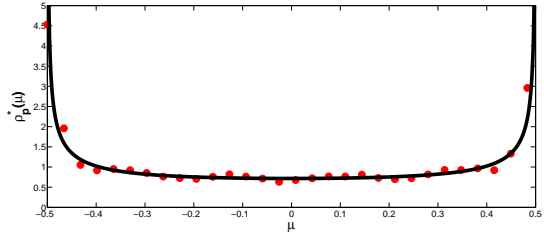}
\caption{(Color online). Density of shifted transmission eigenvalues $\mu$ for
$N=6$ and $p=1$ (theory vs. numerics) for the shot noise case.}
\label{fanoB}
\end{center}
\end{figure}

In fig. \ref{fanoB} we plot the theoretical density of shifted transmission eigenvalues together with Montecarlo simulations for $N=6$ and $p=1$.

\subsection{Support on $(-1/2,-L]\cup [L,1/2)$}\label{FanoDisjoint}

For large negative values of $p$, we envisage a form for the charge density as in fig. 
\ref{sketchfano} (rightmost panel), i.e.
on a disconnected support
with two connected and symmetric components. This is because for large negative $p$,
the external potential $p{\mu}^2$ in \eqref{Ceq} tends to push the charges to
the two extreme edges of the box $0$ and $1$, creating an empty space in the middle.
Since we expect to have a disconnected support, we cannot directly use the single support Tricomi
solution \eqref{formulatricomi}. We need to proceed differently.

We start by recasting eq. \eqref{TricomiQ}  
in the following form:
\begin{align}
\label{prima} p\mu &= \int_{-1/2}^{-L}\frac{\varrho_p^\star({\mu}^\prime)}{\mu-{\mu}^\prime}d{\mu}^\prime+\mathrm{Pr}\int_{L}^{1/2}\frac{\varrho_p^\star({\mu}^\prime)}{\mu-{\mu}^\prime}
d{\mu}^\prime &\quad \mu>0\\
p\mu &= \mathrm{Pr}\int_{-1/2}^{-L}\frac{\varrho_p^\star({\mu}^\prime)}{\mu-{\mu}^\prime}d{\mu}^\prime+\int_{L}^{1/2}\frac{\varrho_p^\star({\mu}^\prime)}{\mu-{\mu}^\prime}d{\mu}^\prime &\quad \mu<0
\end{align}

\begin{figure}[htb]
\begin{center}
\includegraphics[bb =-385   122   998   668,totalheight=0.14\textheight]{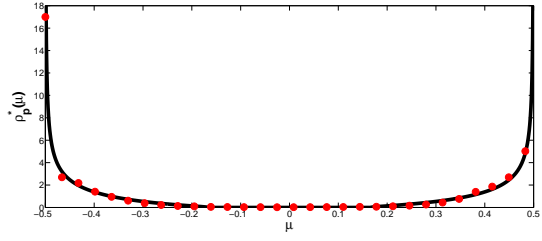}
\caption{(Color online). Density of shifted transmission eigenvalues $\mu$ for
$N=4$ and $p=-9$ (theory vs. numerics) for the shot noise case.}
\label{fanoC}
\end{center}
\end{figure}

In the rhs of \eqref{prima} (first integral) we make the change of variables ${\mu}^\prime\to -{\mu}^\prime$, getting:
\begin{align}
\label{prima2} p\mu &= \int_{L}^{1/2}\frac{\varrho_p^\star(-{\mu}^\prime)}{\mu+{\mu}^\prime}d{\mu}^\prime+\mathrm{Pr}\int_{L}^{1/2}\frac{\varrho_p^\star({\mu}^\prime)}{\mu-{\mu}^\prime}d{\mu}^\prime
\end{align}
Exploiting the symmetry $\varrho_p^\star(\mu)=\varrho_p^\star(-\mu)$, we get:
\begin{align}
\label{prima3} 
\nonumber p\mu &= \mathrm{Pr}\int_{L}^{1/2}d{\mu}^\prime\varrho_p^\star({\mu}^\prime)\left[\frac{1}{\mu+{\mu}^\prime}+\frac{1}{\mu-{\mu}^\prime}\right]\\
&=2\mu\ \mathrm{Pr}\int_L^{1/2}d{\mu}^\prime\frac{\varrho_p^\star({\mu}^\prime)}{{\mu}^2-{\mu}^{\prime 2}}
\end{align}
Making a further change of variables ${\mu}^2=y,{\mu}^{\prime 2}=y^\prime$ we get eventually a Tricomi equation for $\tilde{\varrho}_p(\mu)=\varrho_p^\star(\sqrt{\mu})/\sqrt{\mu}$ as:
\begin{align}
\label{prima4} p &= \mathrm{Pr}\int_{L^2}^{1/4}d y^\prime\frac{\tilde{\varrho}_p(y^\prime)}{y-y^\prime}
\end{align}
Solving \eqref{prima4} by the standard one support solution \eqref{formulatricomi}
and converting back to 
$\varrho_p^\star(\mu)$ we get:
\begin{equation}
\varrho_p^\star(\mu)=\frac{|p\mu|(a_2-{\mu}^2)}{\pi\sqrt{(1/4-{\mu}^2)({\mu}^2-L^2)}}
\end{equation}
where $a_2$ is an arbitrary constant, fixed by the condition $\varrho_p^\star(\pm L)=0$ (the density is vanishing at 
the edge points). This gives $a_2=L^2$.
Imposing the normalization condition, we get $L^2=1/4-2/|p|$.
The condition that $L\ge 0$ implies that this solution is valid
when $p\le -8$. Thus for $p\le -8$, we then get:
\begin{equation}\label{DensityDisjoint}
\varrho_{p}^\star(\mu)=\frac{|p\mu|\sqrt{{\mu}^2-1/4+2/|p|}}{\pi\sqrt{1/4-{\mu}^2}}
\end{equation}
Note that, when $p\to -8$ from below, the equilibrium solution \eqref{DensityDisjoint} smoothly 
matches the solution \eqref{SemicircleAfterConditions1} in the intermediate regime.
The action is readily evaluated from \eqref{spQsimplified} as:
\begin{align}
\nonumber S[\varrho_p^\star] &=p\int_L^{1/2}d\mu {\mu}^2\varrho_{p}^\star(\mu)
-2\int_L^{1/2}d\mu\varrho_p^\star(\mu)\log \mu\\
&=\frac{3}{4}+\frac{1}{2}\log(|p|)-\frac{|p|}{4} +\frac{1}{2}\log 2
\end{align}

The corresponding $J_Q(p)$ is given by:
\begin{equation}\label{JQPdisjoint}
  J_Q(p)=\frac{3}{4}+\frac{1}{2}\log\left(\frac{|p|}{8}\right)-\frac{|p|}{4}
\end{equation}
from which the rate function $\Psi_Q(x)$ can be easily derived:
\begin{widetext}
\begin{equation}
\Psi_Q(x)=\max_p[-xp+J_Q(p)]=\frac{1}{4}-2\log 2 -\frac{1}{2}\log\left(\frac{1}{4}-x\right)\qquad 3/16\leq x\leq 1/4
\end{equation}
\end{widetext}

In fig. \ref{fanoC} we plot the theoretical density of shifted transmission eigenvalues together with Montecarlo simulations for $N=4$ and $p=-9$.

\subsection{Final results for the shot noise case}
To summarize, the density of the
shifted eigenvalues $\{\mu_i\}$ (solution of the saddle point
equation \eqref{TricomiQ}) has the following form:
\begin{widetext}
\begin{equation}\label{DensitySummaryFano}
  \varrho_p^\star(\mu)=
  \begin{cases}
  \frac{p}{\pi}\sqrt{\frac{2}{p}-{\mu}^2} &
  \qquad -\sqrt{\frac{2}{p}}\leq \mu\leq \sqrt{\frac{2}{p}} \qquad p\geq 8\\
\frac{p}{\pi\sqrt{1/4-{\mu}^2}}\left[\frac{8+p}{8p}-{\mu}^2\right] &
  \qquad -1/2\leq \mu\leq 1/2 \qquad -8\leq p\leq 8\\
 \frac{|p\mu|\sqrt{{\mu}^2-1/4+2/|p|}}{\pi\sqrt{1/4-{\mu}^2}} &
  \qquad -1/2\leq \mu\leq -\sqrt{1/4-2/|p|} \vee \sqrt{1/4-2/|p|}\leq \mu\leq 1/2  \qquad p\leq
  -8
  \end{cases}
\end{equation}
\end{widetext}
One may easily check that $\varrho_p^\star(\mu)$ is continuous at $p=\pm 8$,
but develops two phase transitions characterized by different
supports.

The saddle-point action \eqref{spQsimplified} is given by:
\begin{equation}\label{Spstar Fano summary}
  S[\varrho_p^\star]=
  \begin{cases}
  \frac{3}{4}+\frac{1}{2}\log 2+\frac{1}{2}\log p &
 p>8\\
\frac{p}{8}-\frac{p^2}{256}+2\log 2 &
   -8\leq p\leq 8\\
   \frac{3}{4}+\frac{1}{2}\log(|p|)-\frac{|p|}{4} +\frac{1}{2}\log 2 &
  p\leq -8
  \end{cases}
\end{equation}
which is again continuous at $p=\pm 8$.

From \eqref{AsymptoticDecay}, the expression of $J_Q(p)=S[\varrho_p^\star]-S_0[\varrho_0^\star]$ is:
\begin{equation}\label{JQdiPsummary}
  J_Q(p)=
  \begin{cases}
\frac{3}{4}+\frac{1}{2}\log\left(\frac{p}{8}\right) &
 p\geq 8\\
\frac{-p^2}{256}+\frac{p}{8} &
   -8\leq p\leq 8\\
  \frac{3}{4}+\frac{1}{2}\log\left(\frac{|p|}{8}\right)-\frac{|p|}{4} &
  p\leq -8
  \end{cases}
\end{equation}
from which one can derive (in complete analogy with the
conductance case) the rate function for the auxiliary quantity
$Q$:
\begin{equation}\label{PsiQdiPsummary}
  \Psi_Q(x)=
  \begin{cases}
\frac{1}{4}-2\log 2-\frac{1}{2}\log x &
 0\leq x\leq 1/16\\
64\left(x-\frac{1}{8}\right)^2 &
  1/16\leq x\leq 3/16\\
  \frac{1}{4}-2\log 2-\frac{1}{2}\log\left(\frac{1}{4}-x\right) &
  3/16\le x\le 1/4
  \end{cases}
\end{equation}
and from the relation $\Psi_P(x)=\Psi_Q(1/4-x)$ one readily
obtains the rate function for the shot noise in \eqref{PsiPsummary}.

\section{Distribution of moments $\mathcal{T}_n$ for integer $n$}\label{distmom}

In this section, we deal with the more general case of integer
moments $\mathcal{T}_n=\sum_{i=1}^N T_i^n$, in particular focussing on the case $n=2$. The conductance is
exactly given by $\mathcal{T}_1$ while the shot noise is
$\mathcal{T}_1-\mathcal{T}_2$. While we could use the general method outlined in Section \ref{summarycoulomb}
with the choice $a(T)= T^n$, as was done for the conductance case ($n=1$), it turns out
that one can obtain the same final results by using a short-cut which combines, in one step,
the saddle point evaluation in \eqref{inteq1} and the maximization
of the Legendre transform in \eqref{legendre1}. Of course, both methods finally
yield the same results, but this shortcut explicitly avoids any Laplace inversion. Here, we illustrate the short-cut method for the case
$a(T)=T^n$, but it can also be used for other linear statistics.  

The distribution of the moments
$\mathcal{P}_{\mathcal{T}_n}(\mathcal{T}_n=Nt,N)$ is given by:
\begin{widetext}
\begin{equation}\label{DistributionTn}
\mathcal{P}_{\mathcal{T}_n}(\mathcal{T}_n=Nt,N)=A_N\int_0^1\cdots\int_0^1
dT_1\cdots dT_N \exp\left(\frac{\beta}{2}\sum_{j\neq
k}\log|T_j-T_k|+\left(\frac{\beta}{2}-1\right)\sum_{i=1}^N \log
T_i\right)\delta\left(\sum_{i=1}^N T_i^n-Nt\right)
\end{equation}
\end{widetext}
The short-cut consists in replacing the delta function by its integral representation:
$\delta(x)= \int \frac{dp}{2\pi} e^{p x}$ where the integral runs in
the complex $p$ plane. The rest is as before, namely that in the large $N$ limit, one
replaces
the multiple integral by a functional integral introducing
a continuous charge density $\varrho_p(x)$. This gives 
\begin{equation}\label{MultipleToFunctionalTn}
 \mathcal{P}_{\mathcal{T}_n}(\mathcal{T}_n=Nt,N)=
  A_N\int\frac{dp}{2\pi}\mathcal{D}[\varrho]e^{-\frac{\beta}{2} N^2 S[\varrho_p]}
\end{equation}
where the action is given by:
\begin{widetext}
\begin{align}\label{Action Spn}
  S[\varrho_p] =p\left(\int_0^1 \varrho_p(x)x^n dx-t\right)-\int_0^1\int_0^1 dx dx^\prime
  \varrho_p(x)\varrho_p(x^\prime)\log|x-x^\prime|
  +C\left[\int_0^1 \varrho_p(x)dx-1\right]
\end{align}
\end{widetext}
where the rhs of \eqref{MultipleToFunctionalTn} is now extremized with respect to both $\varrho(x)$
and $p$. Notice that here we have already performed the inverse Laplace transform of \eqref{ratio1}.
Hence the two methods are exactly identical. 

Extremizing the action gives the following saddle point
equations:
\begin{align}
 \label{TricomiTn} p x^{n}+C &=2\int_0^1~dx^\prime\varrho_p^\star(x^\prime)\log|x-x^\prime|\\
\label{Tricomi t e p} t &=\int_0^1 ~dx x^n \varrho_p^\star(x) 
\end{align}
which in turn determine $p$ as a function of $t$.

Multiplying \eqref{TricomiTn} by $\varrho_p^\star(x)$ and integrating over $x$, the action at
the saddle point can be rewritten in 
the more compact form:
\begin{equation}\label{Action Spn Improved}
  S[\varrho_p^\star]=-\frac{1}{2}\left[pt+C\right]
\end{equation}
where, as before, the constant $C$ has to be determined from \eqref{TricomiTn} by using
a suitable value of $x$ which is included in the support of the solution. For large $N$,
\eqref{MultipleToFunctionalTn} gives
\begin{align}
\nonumber\mathcal{P}_{\mathcal{T}_n}(\mathcal{T}_n=Nt,N) &\approx \exp\left[-\frac{\beta}{2} N^2 \{S[\varrho_p^\star]-S[\varrho_0^\star]\}\right]=\\
&=\exp\left[-\frac{\beta}{2}N^2\Psi_{\mathcal{T}_n}(t)\right]
\label{largeNaction}
\end{align}
where the rate function is given by
\begin{equation}
\Psi_{\mathcal{T}_n}(t)=S[\varrho_p^\star]- S[\varrho_0^\star]=S[\varrho_p^\star]-2\log 2
\label{ldfim1}
\end{equation}
having used again $S[\varrho_0^\star]=2\log 2$ from \eqref{action0}.

Upon differentiation of \eqref{TricomiTn}, we obtain the Tricomi equation for $\varrho_p^\star(x)$:
\begin{equation}
\label{singular}
 n\, \frac{p}{2}\, x^{n-1}=\mathrm{Pr}\int_0^1 \frac{\varrho_p^\star(x^\prime)dx^\prime}{x-x^\prime}
\end{equation}
to be solved for different supports of $\varrho_p^\star(x)$ as one varies the
argument $t$ and consequently the parameter $p$.  As usual, depending on the 
value of $p$, we need to first anticipate the `type' of the solution, i.e.,
the form of its support and then verify it {\it a posteriori},
as illustrated below.

\subsection{Large $p$: support on $(0,L_p]$}\label{Tn0L1}

Consider first the case when $p$ is very large. Since the external potential
in \eqref{TricomiTn} is of the form $ p\, x^n +C$ which is rather steep
for large $p$, we
anticipate that the charge fluid will be pushed towards the left hard edge at $x=0$.
In this case, the general solution of \eqref{singular} with the
restriction $n\in\mathbb{N}$ is given by:
\begin{align}\label{generalTn}
\nonumber \varrho_{p}^\star(x) &=-\frac{1}{\pi^2\sqrt{x(L_p-x)}} \times\\
 &\times\left[\frac{1}{2}\mathrm{Pr}\int_0^{L_p}\frac{\sqrt{x^\prime 
(L_p-x^\prime)}\, n\,p\, (x^{\prime})^{n-1}\, d x^\prime}{x-x^\prime}
 +C_1\right]
\end{align}
where $C_1$ is an arbitrary constant.
Evaluating the principal value integral yields
\begin{align}
 \nonumber &\varrho_{p}^\star(x) =-\frac{1}{\pi^2\sqrt{x(L_p-x)}}\times\\ 
 &\times\left[-\frac{np 
L_p^n\sqrt{\pi}\Gamma(n-1/2)}{4\Gamma(n+1)}
~_2 F_1\left(1,-n;3/2-n;\frac{x}{L_p}\right)+C_1\right]
\end{align}
where $_2 F_1(a,b;c;x)$ is a
hypergeometric function, defined by the series:
\begin{equation}
_2 F_1(a,b;c;x)=1+\frac{ab}{c}x+\frac{a(a+1)b(b+1)}{c(c+1)}\frac{x^2}{2!}+\cdots
\end{equation}

Determining the constant $C_1$ by the requirement that $\varrho_{p}^\star(L_p)=0$, we obtain:
\begin{align}\label{vr}
 \nonumber \varrho_{p}^\star(x) &=\frac{1}{\pi (2n-1)\sqrt{x(L_p-x)}}\times\\
 &\times\left[
~_2 F_1(1,-n;3/2-n;x/L_p)+2n-1\right]
\end{align}
The edge point $L_p$ is finally determined by the normalization requirement $\int_0^{L_p}\varrho_{p}^\star(x)dx=1$,
yielding after some elementary algebra:
\begin{equation}\label{Lp}
 L_p=\left(2\sqrt{\pi}\Gamma(n)/(p\Gamma(n+1/2))\right)^{1/n}
\end{equation}


Imposing now \eqref{Tricomi t e p} as:
\begin{widetext}
\begin{equation}
 \frac{1}{\pi(2n-1)}\left[\int_0^{L_p}
\frac{x^n}{\sqrt{x(L_p-x)}}\left[2n-1+~_2 F_1\left(1,-n;3/2-n;x/L_p\right)\right]\right]=t
\end{equation}
\end{widetext}
we obtain a very simple relation between $p$ and $t$:
\begin{equation}\label{pt}
 p=\frac{1}{tn}
\end{equation}
Armed with \eqref{pt} and \eqref{vr}, we can now evaluate the action
\eqref{Action Spn Improved} eliminating $p$ as:
\begin{align}\label{Action Spn doubly Improved}
\nonumber S[\varrho_p^\star] &=\frac{1}{2n}+\log\left(\frac{4}{L_p}\right) =\\
&=\frac{1}{2n}+2\log 2+\frac{1}{n}\log\left[\frac{\Gamma(n+1/2)}{2\sqrt{\pi}\Gamma(n+1)}t^{-1}\right]
\end{align}

Equation \eqref{vr} is valid as long as $L_p<1$ 
(the edge point of the support such that $\varrho_p^\star(L_p)=0$).
From \eqref{Lp}, putting $L_p=1$, one thus finds that the solution is valid for $p>p_1^\star$ 
where
\begin{equation}
p_1^\star= \frac{2\sqrt{\pi} \Gamma(n)}{\Gamma(n+1/2)}.
\label{p1star}
\end{equation}
Consequently, from \eqref{pt}, it follows that the solution is valid 
for $t<t_1^\star$, where
\begin{equation}
 t_1^\star=\frac{\Gamma(n+1/2)}{2\sqrt{\pi}\Gamma(n+1)}
\label{t1star}
\end{equation}

As a check, for $n=1$ we have $t_1^\star=1/4$ from \eqref{t1star}, $L_p=4/p=4t$ from \eqref{Lp}
and \eqref{pt} and the 
condition $L_p<1$ 
implies $t<t_1^\star=1/4$ as expected (compare with subsection \ref{conductance0L1}).
For $n=2$, we have $t_1^\star=3/16$ and $p_1^\star= 8/3$. For $n=2$, this regime 
($t<t_1^\star=3/16$ and hence $p>p_1^\star=8/3$) corresponds to
the leftmost panel in Fig. \ref{schema}.

In this region of $t$, the rate function is easily computed as
\begin{align}\label{RateFunction p>p^(+) computed}
\nonumber \Psi_{\mathcal{T}_n}(t) &=S[\varrho^\star]-2\log 2=\\
&=\frac{1}{n}\left[\frac{1}{2}+
\log\left(\frac{\Gamma(n+1/2)}{2\sqrt{\pi}\Gamma(n+1)}\right)-\log t\right]
\end{align}
Combining \eqref{RateFunction p>p^(+) computed} with \eqref{largeNaction}, one obtains as a new result the precise left tail asymptotics for the $n$-th integer moment distribution:
\begin{equation}\label{lefttail}
\mathcal{P}_{\mathcal{T}_n}(\mathcal{T}_n=Nt,N)\approx \exp\left(\frac{\beta N^2}{2n}\log t\right)=t^{\frac{\beta N^2}{2n}}
\end{equation}

\begin{figure}[htb]
\begin{center}
\includegraphics[bb =87   262   507   578,totalheight=0.14\textheight]{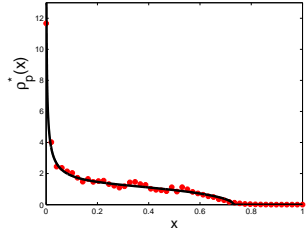}
\caption{(Color online). Density of eigenvalues $x$ for
$N=5$ and $p=5$ (theory vs. numerics) for $n=2$.}\label{mom1}
\end{center}
\end{figure}

In fig. \ref{mom1} we plot the theoretical density of eigenvalues for the $n=2$ case together with Montecarlo simulations with $N=5$ and $p=5$.

\subsection{Intermediate $p$: support on $(0,1)$}\label{1s}

We now look for the solution of 
\begin{equation}\label{tricmn}
 n\frac{p}{2} x^{n-1}=\mathrm{Pr}\int_0^1\frac{\varrho_p^\star(x^\prime)}{x-x^\prime}d x^\prime
\end{equation}
with a nonzero support over the full allowed range $[0,1]$. Using the general solution
in \eqref{formulatricomi} with the choice $a=0$ and $b=1$ we get
\begin{equation}\label{varrhoTn01}
\varrho_{p}^\star(x)=\frac{np\Gamma(n-1/2)}{4\Gamma(n+1)\pi^{3/2}}
\left[\frac{~_2 F_1(1,-n;3/2-n;x)-C_n}{\sqrt{x(1-x)}}\right]
\end{equation}
where $C_n$ is an arbitrary constant to be fixed by $\int_0^1 \varrho_{p}^\star(x)dx=1 $, yielding:
\begin{equation}\label{constantC}
 C_n=-\frac{4\sqrt{\pi}\Gamma(n)}{p\Gamma(n-1/2)}
\end{equation}
so eventually:

\begin{equation}\label{varrhoTn02}
\varrho_{p}^\star(x)=\frac{p\Gamma(n-1/2)}{4\Gamma(n)\pi^{3/2}}\left[\frac{~_2 
F_1(1,-n;3/2-n;x)+\frac{4\sqrt{\pi}\Gamma(n)}{p\Gamma(n-1/2)}}{\sqrt{x(1-x)}}\right]
\end{equation}
Next, we need to impose the condition \eqref{Tricomi t e p}:
\begin{equation}\label{Tricomi t and p imposed}
 t=\int_0^1 dx~x^n \varrho_{p}^\star(x)
\end{equation}
leading to:
\begin{equation}\label{tpmoment01}
 t=\frac{\Gamma(n+1/2)}{\sqrt{\pi}\Gamma(n+1)}\left[1+\frac{p\Gamma(n-1/2)}{8\sqrt{\pi}\Gamma(n)}(1-2n)\right]
\end{equation}
From \eqref{tpmoment01} we can derive the relation between $p$ and $t$ as:
\begin{equation}\label{ptmom}
 p=a_n-b_n t
\end{equation}
where:
\begin{align}
 a_n &=\frac{4\sqrt{\pi}\Gamma(n)}{\Gamma(n+1/2)}\\
b_n &=\frac{4\pi\Gamma(n)\Gamma(n+1)}{[\Gamma(n+1/2)]^2}
\end{align}

Inserting \eqref{ptmom} and \eqref{varrhoTn02} into \eqref{Action Spn Improved}, we obtain, after
a few steps of algebra, the 
action:
\begin{equation}
 S[\varrho_p^\star]=\frac{b_n}{2}\left[t-\frac{\Gamma(n+1/2)}{\sqrt{\pi}\Gamma(n+1)}\right]^2+\log 4
\end{equation}

Next we need to determine the range of validity of this solution.
This is obtained simply by the fact that the density $\varrho_p^\star(x)$ in \eqref{varrhoTn01}
must be positive. Let us first rewrite the solution \eqref{varrhoTn01} as
\begin{equation}
\varrho_p^\star(x)= \frac{1}{\pi}\, \frac{A_p(x)}{\sqrt{x(1-x)}}
\label{sol1}
\end{equation}
where 
\begin{equation}
A_p(x)= 1+ \frac{p\Gamma(n-1/2)}{4\sqrt{\pi} \Gamma(n)}\, _2 F_1(1,-n;3/2-n;x).
\label{amp1}
\end{equation}
To ensure $\varrho_p^\star(x)\ge 0$, we have to just ensure that $A_p(x)\ge 0$ in \eqref{amp1}.
How does $A_p(x)$ vary as a function of $x$ in $x\in [0,1]$? It can be easily seen that
this function has a global minimum at some intermediate value $0<x^\star<1$.
To ensure its positivity, we then have to ensure that $A_p(x^\star)\ge 0$.
This will be true only for a range of values of $p$, i.e., when
$p_2^\star \le p\le p_1^\star$ (where $p_1^\star$ is precisely the lower
edge of the validity of regime I in the previous subsection and is given in \eqref{p1star}).
Consequently, using \eqref{ptmom}, this sets a $t$ range $t_1^\star\le t \le t_2^\star$
for the validity of this regime II, where $t_1^\star$ is given in \eqref{t1star}.
Now, for arbitrary $n$, $p_2^\star$ and consequently $t_2^\star$ have rather
complicated expressions which we do not detail here. But for $n=2$, their expressions
are rather simple and we get
\begin{equation}
p_2^\star= -\frac{16}{3};\quad {\rm and}\quad t_2^\star = \frac{3}{4}
\label{p2t2star}
\end{equation}
This then defines regime II with a full support over $[0,1]$, namely
$-16/3\le p\le 8/3$ and consequently $3/16\le t\le 3/4$, is shown
as the second (from the left) region in fig. \ref{schema}.

Thus in this regime II where $t_1^\star \le t_2^\star$, the rate function, for arbitrary $n$, has a
quadratic form:
\begin{equation}
\Psi_{\mathcal{T}_n}(t)=S[\varrho_p^\star]-\log 4=
\frac{b_n}{2}\left[t-\frac{\Gamma(n+1/2)}{\sqrt{\pi}\Gamma(n+1)}\right]^2
\end{equation}
thus implying \eqref{DecayPTn}:
\begin{equation}\label{DecayPTnComputed}
\mathcal{P}_{\mathcal{T}_n}(\mathcal{T}_n,N)\approx
\exp\left[-\frac{\beta}{2}N^2 \frac{b_n}{2}\left[\frac{\mathcal{T}_n}{N}-\frac{\Gamma(n+1/2)}{\sqrt{\pi}\Gamma(n+1)}\right]^2\right]
\end{equation}

\begin{figure}[htb]
\begin{center}
\includegraphics[bb =87   262   507   578,totalheight=0.14\textheight]{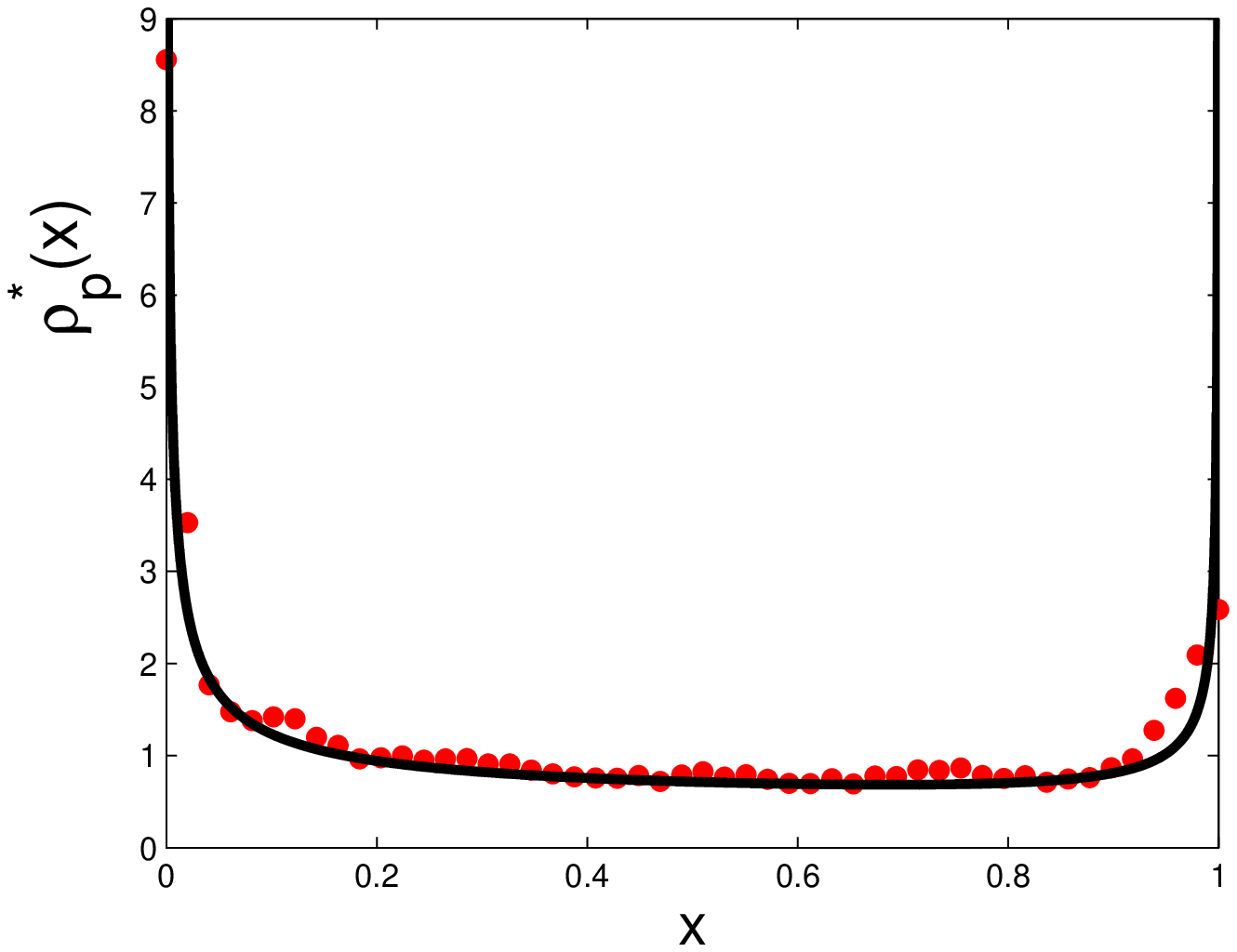}
\caption{(Color online). Density of eigenvalues $x$ for
$N=6$ and $p=1$ (theory vs. numerics) for $n=2$.}\label{mom2}
\end{center}
\end{figure}

In fig. \ref{mom2} we plot the theoretical density of eigenvalues for the $n=2$ case together with Montecarlo simulations with $N=6$ and $p=1$.

From the Gaussian shape in \eqref{DecayPTnComputed}, the mean and
variance of $\mathcal{T}_n$ can be read off very easily:
\begin{align}
\label{meanmoment}\langle\mathcal{T}_n\rangle &=\frac{N\Gamma(n+1/2)}{\sqrt{\pi}\Gamma(n+1)}\\
\label{varmoment}\mathrm{var}(\mathcal{T}_n) &=\frac{2}{\beta b_n}=\frac{[\Gamma(n+1/2)]^2}{2\beta\pi\Gamma(n)\Gamma(n+1)}
 \end{align}
\begin{figure}[htb]
\begin{center}
\includegraphics[bb =91 3 322 146,totalheight=0.20\textheight]{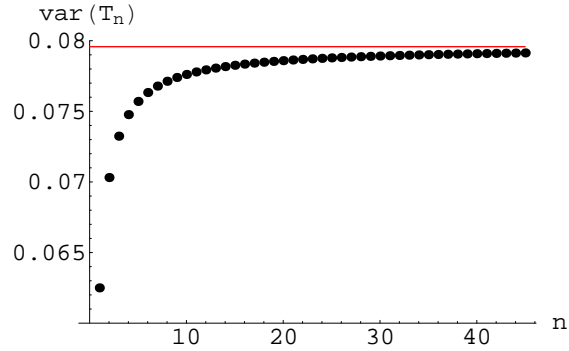}
\caption{(Color online). $\mathrm{var}(\mathcal{T}_n)$ as a function of
$n\in\mathbb{N}$ \eqref{varmoment} for $\beta=2$. In red the
asymptotic value $v^\star=1/2\pi\beta\approx 0.07957$ \eqref{Asymptotic vstar}.\label{vartnfig}}
\end{center}
\end{figure}
Novaes \cite{novaes} recently computed the average of integer moments and obtained
the following expression for arbitray $n\in\mathbb{N}$:
\begin{equation}\label{novaesformula}
\langle\mathcal{T}_n\rangle=N\binom{2n}{n}4^{-n}
\end{equation}
Using elementary properties of Gamma functions, it is easy to show that formulae \eqref{meanmoment} and \eqref{novaesformula} do indeed 
coincide,
a fact not completely apparent at first sight.

Conversely, the
exact expression for the large $N$ variance \eqref{varmoment} is new\footnote{Novaes has recently
computed the variance for any finite number of open channels $N_1$ and
$N_2$ \cite{novaes}: however, extracting the asymptotics from his formula
does not appear to be easy. }, since the general integral
in Beenakker's formula \cite{beenakkerPRL} does
not appear easy to carry out explicitly. Obviously, \eqref{varmoment}
agrees with the known result \cite{beenakkerPRL}
for the conductance $\mathrm{var}(G)=1/8\beta$ for $n=1$.

From \eqref{varmoment}, it is easy to extract the asymptotic value:
\begin{equation}\label{Asymptotic vstar}
  v^\star=\lim_{n\to\infty}\mathrm{var}(\mathcal{T}_n)=\frac{1}{2\beta\pi}
\end{equation}
which is plotted in fig. \ref{vartnfig} for $\beta=2$ together
with \eqref{varmoment}.

For general $n$, as one increases the value of $t$, so far we have seen two regimes: regime I ($0\le t\le
t_1^\star$)
with support over $[0,L_p]$ and then regime II ($t_1^\star\le t \le t_2^\star$) with
support over the full range $[0,1]$. What happens
when $t$ increases beyond $t_2^\star$? For arbitrary $n$, the analysis becomes rather cumbersome.
So from now we restrict ourselves only to the case $n=2$ (which turns out already to be rather
nontrivial). But at least for $n=2$ we are able to obtain a full picture and in the
two subsections below we show that apart from regime I and regime II already discussed
above, two further regimes appear as one increases $t$ beyond $t_2^\star$: 
regime III (for $t_2^\star\le t\le t_3^\star$) where the solution
has a disconnected support with two connected components discussed in subsection \ref{2s} and
regime IV (for $t_3^\star\le t\le 1$) where the solution again has
a single support but on the other side of the box over $[M_p,1]$.
Since the solution in regime IV is simpler (single support), we will
first discuss this case in the next subsection \ref{3s} and finally
the more involved case of regime III (with a disconnected support) will
be discussed in subsection \ref{2s}.

\subsection{Support on $[M_p,1)$ ($n=2$): regime IV}\label{3s}

Focussing on the $n=2$ case, we now look for a solution
of \eqref{singular} with a single support $[M_p,1]$
where $M_p$ is yet to be determined. Using the general
single support Tricomi solution \eqref{formulatricomi} choosing $a=M_p$ and $b=1$,
one obtains the following explicit solution
\begin{widetext}
\begin{equation}\label{densitychargesMp1}
 \varrho_{p}^\star(x)=-\frac{1}{\pi^2\sqrt{(x-M_p)(1-x)}}\left[p~\mathrm{Pr}
\int_{M_p}^{1}\frac{\sqrt{(1-x^\prime)(x^\prime-M_p)}}{x-x^\prime}\, x^\prime\, d x^\prime+ D\right]
\end{equation}
\end{widetext}
where $D$ is an arbitrary constant.
Evaluating the principal value integral in \eqref{densitychargesMp1} 
and imposing $\varrho_{p}^\star(M_p)=0$ we obtain
\begin{equation}\label{densitychargesMp1part3}
 \varrho_{p}^\star(x)=\frac{-p(2x+M_p-1)}{2\pi}\sqrt{\frac{x-M_p}{1-x}}
\end{equation}
The lower edge $M_p$ is determined by the normalization condition $\int_{M_p}^1 \varrho_{p}^\star(x)dx=1$,
yielding a quadratic equation for $M_p$: $p(M_p-1)(1+3M_p)=8$ with two roots
$M_p= (1\pm 2\sqrt{1+6/p})/3$. Noting that when $p\to -\infty$, it follows from
physical consideration that the charge density must be pushed to its rightmost limit indicating
that $M_p\to 1$ as $p\to -\infty$. This condition forces us to choose the correct root as
\begin{equation}\label{Mp}
 M_p=\frac{1}{3}\,\left(1+2\sqrt{1+\frac{6}{p}}\right).
\end{equation}
The condition $0\leq M_p\leq 1$ implies for $p$ the condition $p\le p_3^\star=-6$.
The relation between $p$ and $t$ is then obtained using \eqref{Tricomi t e p}, resulting in the condition:
\begin{equation}\label{unwieldly}
 \frac{15+27 M_p+13 M_p^2+9 M_p^3}{16(3 M_p+1)}=t
\end{equation}
where $M_p$ is expressed as a function of $p$ in \eqref{Mp}. 
The solution of \eqref{unwieldly} is quite cumbersome to write down explicitly, 
but is in principle feasible. Note that when $p\to -6$ from below, $M_p\to 1/3$ from
\eqref{Mp} and consequently from \eqref{unwieldly}, $t\to t_3^\star=29/36$ from above.
In other words, the solution \eqref{densitychargesMp1part3} is valid in
regime IV defined by
\begin{equation}
p\le p_3^\star=-6;\quad  {\rm consequently}, \,\quad t_3^\star=\frac{29}{36} \le t \le 1
\label{regime4}
\end{equation}
This regime IV is shown in the extreme right part of Fig. \ref{schema}.

Once $p$ has been determined has a function of $t$ from
\eqref{Mp} and \eqref{unwieldly} and substituted into the density \eqref{densitychargesMp1part3}, 
the action and the rate function can be computed from \eqref{Action Spn Improved}
and \eqref{ldfim1}, by evaluating 
numerically the corresponding integrals. We omit these details
here.

\begin{figure}[htb]
\begin{center}
\includegraphics[bb =87   262   507   578,totalheight=0.14\textheight]{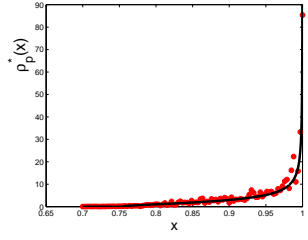}
\caption{(Color online). Density of eigenvalues $x$ for
$N=4$ and $p=-10$ (theory vs. numerics) for $n=2$.}\label{mom3}
\end{center}
\end{figure}

In fig. \ref{mom3} we plot the theoretical density of eigenvalues for the $n=2$ case together with Montecarlo simulations with $N=4$ and $p=-10$.

\subsection{Disconnected support ($n=2$): regime III}\label{2s}

For $n=2$, it then remains to find the solution of \eqref{singular} in the narrow band $t_2^\star=3/4\le t\le 
t_3^\star=29/36$ or equivalently for $p_3^\star=-6 \le p \le p_2^\star=-16/3$ (see Fig. \ref{schema}).
This is the regime III. Let us first try to anticipate what the solution
may look like in this regime. For this, let us consider the two regimes namely regime II
and regime IV respectively to the left and right of regime III. 

Consider first regime II
($t_1^\star=3/16 \le t\le t_2^\star=3/4$). In this regime the solution $\varrho_p^\star(x)$ has a single support
over the full range $x\in [0,1]$ given in \eqref{varrhoTn02}, which for $n=2$ (using
the special value of the hypergeometric function) simply reads, with $p_2^\star=-16/3\le p \le p_1^\star=8/3$,
\begin{equation}
\varrho_p^\star(x) = \frac{1}{\pi \sqrt{x(1-x)}}\left[1+\frac{p}{8}(1+4x-8x^2)\right].
\label{n2sol2}
\end{equation}
Now, when $t$ tends to the maximum allowed value in regime II, namely, $t\to t_2^\star=3/4$ from below
or equivalently $p\to p_2^\star=-16/3$ from above, the solution in \eqref{n2sol2} 
tends to
\begin{equation}
\varrho_{-16/3}^\star(x) = \frac{16}{3\pi \sqrt{x(1-x)}}\, \left[x-\frac{1}{4}\right]^2
\label{n2solc2}
\end{equation}
with a quadratic minimum at $x=1/4$ where the density vanishes. This is just the edge of
regime II. If $t$ increases slightly beyond $t_2^\star=3/4$, this single support solution
is no longer valid. However, it gives the hint that for $t> t_2^\star=3/4$, the charge
density must separate into two disjoint components, one on the left side over $[0,l_1]$ and one on the
right side over $[l_2,1]$ with an empty stretch $[l_1,l_2]$ separating them. This empty stretch must increase
as one increases $t$ beyond $t_2^\star=3/4$ in this regime III. Indeed, as $t$ increases
further, the left support $[0,l_1]$ must shrink in size and the right support $[l_2,1]$ must
increase in size and finally when $t$ hits the value $t_3^\star=29/36$, $l_1$ must shrink to $0$
and $l_2$ must approach $M_p=1/3$ and then one arrives in regime IV discussed in the
previous subsection. At exactly $t=t_3^\star=29/36$ or equivalently at $p=p_3^\star=-6$, 
the solution (border of regime IV) can be read off \eqref{densitychargesMp1part3} with $M_p=1/3$
\begin{equation}
\varrho_{-6}^\star(x)= \frac{6}{\pi \sqrt{1-x}}\, \left(x-\frac{1}{3}\right)^{3/2}.
\label{n2solc4}
\end{equation}

Thus, the solution in regime III, namely for $p_3^\star=-6\le p \le p_2^\star =-16/3$ must interpolate
between the solutions given in \eqref{n2solc2} and in \eqref{n2solc4} valid respectively
at the two edges of regime III and have a disconnected support with two connected components over $[0,l_1]$ and $[l_2,1]$.  
We were able to find this solution explicitly. Its derivation is outlined
in the Appendix. The result reads
\begin{equation}
\varrho_p^\star(x)= \frac{-p}{\pi \sqrt{x(1-x)}}\,\sqrt{(x-l_1)(x-l_2)^3}
\label{exsol1}
\end{equation}
which is valid for all $0\le x\le l_1$ and $l_2\le x\le 1$ and the two edges $l_1$ and $l_2$
are given by
\begin{eqnarray} 
l_1 & =& \frac{1}{4}\left(1-3 \sqrt{1+\frac{16}{3p}}\right) \label{l1} \\
l_2 &=& \frac{1}{4}\left(1+\sqrt{1+\frac{16}{3p}}\right) \label{l2}
\end{eqnarray}
Note that $l_1$ and $l_2$ are real only if $p\le -16/3$. Furthermore $l_1\ge 0$
only if $p\ge -6$. Thus this solution is valid over the full range $p_3^\star=-6 \le p\le p_2^\star=-16/3$.
This then defines regime III. Note also that the solution \eqref{exsol1} smoothly interpolates
between the solutions in \eqref{n2solc2} (when $p\to -16/3$) and in \eqref{n2solc4} (when $p\to -6$).

The relation between $t$ and $p$ can be obtained substituting \eqref{exsol1} in \eqref{Tricomi t e p}
and performing the integral. Similarly the action and the rate function can be computed
by using the exact density and performing the integrals in \eqref{Action Spn Improved}
and \eqref{ldfim1} by Mathematica, the details 
of which we omit here.
The Montecarlo simulations in this regimes are much harder to obtain due to large fluctuations in sampling from the leftmost residual band $0\leq x\leq l_1$ and a very large $N$
is necessary to achieve a satisfactory picture. Nevertheless, we observed upon increasing $N$ a trend in the equilibrium density which is fully compatible with the analytical 
disconnected-support solution found above.

\subsection{Final results for the $n=2$ moment case}

\begin{figure}[htb]
\begin{center}
\includegraphics[bb =0 0 567 785,totalheight=0.45\textheight]{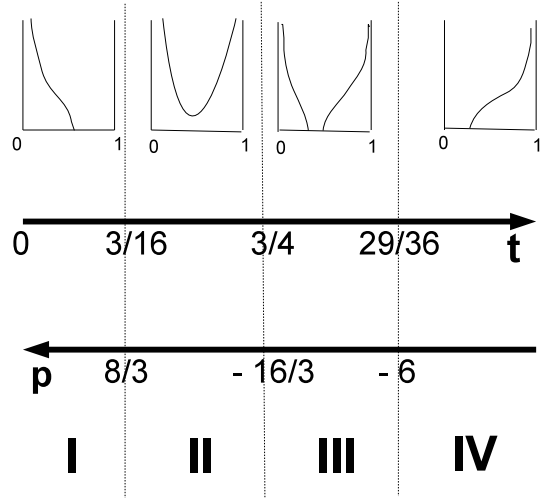}
\caption{Schematic table summarizing the relevant phases in the $t$ and $p$ space for the density of $n=2$ moment.
As $t$ varies in the allowed range $[0,1]$ and consequently the parameter $p$ in $[-\infty, \infty]$,
the fluid density displays 4 different phases as shown in the top panel. These $4$ regimes
are separated by three critical points: $t_1^\star=3/16$ (consequently $p_1^\star=8/3$),
$t_2^\star= 3/4$ ($p_2^\star=-16/3$) and $t_3^\star=29/36$ ($p_3^\star=-6$).
Consequently, the rate function 
$\Psi_{\mathcal{T}_n}(t)$ has 4 different expressions
according to different regions in the $t\in [0,1]$ segment.\label{schema}}
\end{center}
\end{figure}

In fig. \ref{schema} we propose a schematic summary of the different phases of the density of integer moments
for the case $n=2$. As a consequence, the rate function will display three (and not just two, as for conductance and shot noise) non-analytical points corresponding to physical phase transitions in Laplace space.
Starting from high $p$ values, the fluid particles are initially confined towards the left hard edge. Then, when $p$ hits the critical value $p=8/3$, the fluid spreads over the entire support $[0,1]$.
In the narrow region $-6\leq p\leq -16/3$ the density splits over two disjoint (non-symmetric) components of the support, and the leftmost disappears upon hitting the value $p=-6$, leaving the charges leaning against the right hard wall.

\section{Conclusions}

The Coulomb gas analogy, together with recently introduced functional methods, brings about an efficient formalism
for the computation of full probability distributions of observables (valid for a large number of electronic channels in the two leads) in the quantum conductance problem.
Generically, the distribution of any linear statistics of the form $A=\sum_{i=1}^N a(T_i)$, where the $T_i$'s are transmission eigenvalues of the cavity and $a(x)$ is any smooth function,
can be derived within the formalism described in this paper: the problem amounts to finding the equilibrium configuration of an associated 2d charged fluid confined to the segment $[0,1]$
and subject to two competing interactions: the logarithmic repulsion, generated by the Vandermonde term in the jpd \eqref{jpd transmission First}, and a confining potential whose strength is tuned by the Laplace parameter $p$.
Interestingly, this auxiliary Coulomb gas undergoes different phase transitions as the Laplace parameter is varied continuously, and this physical picture is mirrored in the appearance of very weak singularities
in the rate functions of observables at the critical points. Already the leading $N$-term of the free energy of such a gas (the \emph{spherical} contribution) displays non-Gaussian features in the tails,
while the central region obeys the Gaussian law, in agreement with the general Politzer's argument \cite{politzer}; conversely the tails follow
a power-law decay and the junctions of the two (or more, as in the case of integer moments) regimes are continuous but non-analytical points. Note that it is not necessary to develop
a $(1/N)$ expansion of the free energy and look for higher genus terms to appreciate deviations from the Gaussian law.
From the central Gaussian region, one can easily reads off the mean and variance of any linear statistics of interest: this way well known results for e.g. conductance and shot noise are recovered
and new formulas (such as the variance of integer moments and its universal asymptotics) can be derived.  Our results are well-corroborated by Montecarlo simulations both in the real and Laplace space,
as well as by comparison with exact finite $N$ results when available, which convincingly disprove the large-$N$ asymptotic analysis performed in \cite{Kanz}. In summary, the Coulomb gas method (well-suited to large $N$ evaluations) reveals a rich 
thermodynamic behavior for the quantum conductance problem \emph{already at the leading order level}, and we expect that it will enjoy a broad range of applicability.

\begin{acknowledgments}
We are grateful to
C\'{e}line Nadal for helping us with the Monte Carlo simulations.
\end{acknowledgments}

\appendix

\section{Explicit Two Support Solution for $n=2$ in Regime III}

We consider the integral equation \eqref{singular} for $n=2$
\begin{equation}
\label{singularb1}
p x =\mathrm{Pr}\int_0^1 \frac{\varrho_p^\star(x^\prime)dx^\prime}{x-x^\prime}
\end{equation}
and look for a solution that has a disconnected support with two connected components ($[0,l_1]$ and $[l_2,1]$).
This solution is valid in regime III discussed in subsection VI-D.
While a general single support solution of the singular integral equation
can be found using Tricomi's formula given in \eqref{formulatricomi} with $g(x)=px$
and discussed in subsections VI-A, VI-B and and VI-C, it is much more complicated
to obtain an explicit solution with more than one connected component of the support. To find such
a solution, we actually use an alternative method originally used by Brezin et. al.
to find a single support solution of the singular integral equation in the
context of counting of planar diagrams~\cite{brezin}.
This method consists in making a judicious guess for the solution
and then uses the uniqueness properties of analytic functions
in a complex plane to prove that the guess is right.  
Although, for the single support solution 
one does not have to use this route since the explicit general
solution of Tricomi is available (  
the authors of ref.~\cite{brezin} were perhaps unaware of the general
single support explicit solution of Tricomi). Nevertheless, this alternative
method of ~\cite{brezin} can be fruitfully adapted to find a two support solution
(as in our case) in a simpler way (as demonstrated below), where a general 
solution is somewhat difficult to obtain explicitly.

Let us assume that the solution $\varrho_p^\star(x)$ of \eqref{singularb1} has
a disconnected support with two connected components $[0,l_1]$ and $[l_2,1]$ where $l_1$ and $l_2\ge l_1$ are
yet to be determined. Generalizing the route used for single support solution
in ref.~\cite{brezin}, we first introduce an analytic function (without the
principal part in \eqref{singularb1})
\begin{equation}
\label{regularb1}
F(x) =\int_0^{l_1} \frac{\varrho_p^\star(x^\prime)dx^\prime}{x-x^\prime}
+\int_{l_2}^1 \frac{\varrho_p^\star(x^\prime)dx^\prime}{x-x^\prime}
\end{equation}
defined everywhere in the complex $x$ plane outside the two real intervals
$[0,l_1]$ and $[l_2,1]$. This new function $F(x)$ has the following properties:
\begin{enumerate}

\item it is analytic in the complex $x$ plane outside the two cuts $[0,l_1]$ and $[l_2,1]$,

\item  it behaves as $1/x$ when $|x|\to \infty$ since $\int 
\varrho_p^\star(x^\prime)dx^\prime=1$
due to normalization,

\item  it is real for $x$ real outside the two cuts $[0,l_1]$ and $[l_2,1]$,

\item  as one approaches to any point $x$ on the two cuts $[0,l_1]$ and $[l_2,1]$ on the
real axis, $F(x\pm 
i\epsilon)= px \mp i \pi \varrho_p^\star(x)$. This last property follows
from \eqref{singularb1}.

\end{enumerate}

From the general properties of analytic functions in the complex plane it follows
that there is a unique function $F(x)$ which satisfies all the four properties
mentioned above. Thus, if one can make a good guess for $F(x)$ and one verifies
that it satisfies all the above properties, then this $F(x)$ is unique. Knowing $F(x)$,
one can then read off the solution $\varrho_p^\star(x)$ from the $4$-th
property. It then rests to make a good guess for $F(x)$. We try
the following ansatz for $F(x)$ valid everywhere outside the two cuts $[0,l_1]$
and $[l_2,1]$
\begin{equation}
F(x) = px - \frac{p}{\sqrt{x(x-1)}} \sqrt{(x-l_1)(x-l_2)^3}.
\label{guess1}
\end{equation}
This ansatz clearly satisfies the first property. Now, expanding $F(x)$ for large $|x|$
we get
\begin{align}
\nonumber F(x) &\to (-1+l_1+3l_2)p +\\
&+\frac{p}{8}(-3+2l_1 +l_1^2+6l_2-6l_1 l_2-3l_2^2)\,\frac{1}{x}+O(x^{-2}) 
\label{guess2}
\end{align}
Since the second property dictates that $F(x)\to 1/x$, it follows that we must have
\begin{eqnarray}
& l_1+3l_2=1 \label{g1} \\
& (-3+2l_1 + l_1^2+6l_2-6l_1 l_2-3l_2^2)=8 \label{g2}
\end{eqnarray}
Eliminating $l_2$ in \eqref{g2} using \eqref{g1} gives a quadratic equation
for $l_1$, $2l_1^2-l_1-1=6/p$ with two solutions: $l_1= (1\pm 3\sqrt{16/p})/4$.
The correct root is chosen by the fact that when $p\to -6$, $l_1\to 0$ as follows
from the solution in \eqref{n2solc4}. This then uniquely fixes
$l_1$ and $l_2$ given respectively in \eqref{l1} and \eqref{l2}.

The ansatz $F(x)$, with the choices $l_1$ and $l_2$ as in \eqref{l1} and \eqref{l2}
then satisfies the second property. It is easy to check that $F(x)$ satisfies
the third property as well. From the fourth property one then reads off
the unique solution as given in \eqref{exsol1}. This two support solution 
is clearly valid only in the regime III namely for $p_3^\star=-6\le p \le p_2^\star =-16/3$
and it smoothly matches with the solutions of regime II and regime IV respectively
as $p\to -16/3$ and $p\to -6$.

\end{document}